\documentclass[twocolumn]{aa}  
\usepackage{graphicx}
\usepackage{txfonts}
\newcommand{\etal}{et al.}

\newcommand{\gtsim}{\raisebox{-1mm}{$\stackrel{>}{\sim}$}}
\def\arcm{\hbox{$^\prime$}}
\def\arcs{\hbox{$^{\prime\hskip -0.1em\prime}$}}

\begin{document}

   \title{A New Comprehensive 2-D Model of the Point Spread Functions of the XMM-Newton EPIC Telescopes : Spurious Source Suppression and Improved Positional Accuracy}

   \author{A.\,M. Read\inst{1}
           S.\,R. Rosen\inst{1}
           R.\,D. Saxton\inst{2}
           J. Ramirez\inst{3}}

   \offprints{A.\,M. Read}

   \institute{\inst{1}Dept.\ of Physics and Astronomy, Leicester University, Leicester LE1\,7RH, U.K.\\
              \email{amr30@star.le.ac.uk} \\
             \inst{2}XMM SOC, ESAC, Apartado 78, 28691 Villanueva de la Ca\~{n}ada, Madrid, Spain\\
             \inst{3}Leibniz-Institut f\"{u}r Astrophysik Potsdam, An der Sternwarte 16 D-14482 Potsdam, Germany\\
}

   \titlerunning{A New Comprehensive 2-D Model of the Point Spread Functions of the XMM-Newton EPIC Telescopes}
   \authorrunning{A.\,M. Read et al.}

   \date{Received September 15, 1996; accepted March 16, 1997}


  \abstract
{}
{We describe here a new full 2-D parameterization of the PSFs of the
  three XMM-Newton EPIC telescopes as a function of instrument,
  energy, off-axis angle and azimuthal angle, covering the whole
  field-of-view of the three EPIC detectors. It models the general PSF
  envelopes, the primary and secondary spokes, their radial
  dependencies, and the large-scale azimuthal variations.  }
{This PSF model has been constructed via the stacking and centering of
  a large number of bright, but not significantly piled-up point
  sources from the full field-of-view of each EPIC detector, and
  azimuthally filtering the resultant PSF envelopes to form the spoke
  structures and the gross azimuthal shapes observed. }
{This PSF model is available for use within the XMM-Newton Science
  Analysis System via the usage of Current Calibration Files
  XRT$i$\_XPSF\_0011.CCF and later versions. Initial source-searching tests
  showed substantial reductions in the numbers of spurious sources being
  detected in the wings of bright point sources. Furthermore, we have
  uncovered a systematic error in the previous PSF system, affecting
  the entire mission to date, whereby returned source RA and Dec
  values are seen to vary sinusoidally about the true position (amplitude$\approx$0.8\arcs) with source azimuthal
  position.}
{The new PSF system is now available and is seen as a major
  improvement with regard to the detection of spurious sources. The
  new PSF also largely removes the discovered astrometry error and is
  seen to improve the positional accuracy of EPIC. The modular nature
  of the PSF system allows for further refinements in the future.}


   \maketitle

\section{Introduction}

XMM-Newton (Jansen \etal\ 2001), a cornerstone mission of ESA's
Horizon 2000 science program, was designed as an X-ray observatory
able to study cosmic X-ray sources spectroscopically with the highest
possible collecting area in the 0.2$-$10\,keV band. This high
throughput is achieved primarily through the use of 3 highly nested
Wolter type {\rm I} imaging telescopes. The design of the optics was
driven by the need to have the highest possible effective area up to
10\,keV, and in particular at $\sim$7\,keV, where the K lines of
astrophysically significant iron appear. In grazing incidence optics,
the effective area is generally increased by nesting many mirrors
together and packing the front aperture as much as possible. In the
case of the XMM-Newton mirrors, each of the three telescopes contains
a mirror module of focal length 7.5\,m, comprising 58 nested mirror
shells, the axial length of the total mirror being 60\,cm, this shared
equally between the paraboloid and the hyperboloid halves of the
Wolter~{\rm I} configuration. The maximum diameter mirror shell is
70\,cm, and the outer and inner mirror shell thicknesses are 1.07\,mm
and 0.47\,mm respectively (Aschenbach \etal\ 2000).

One of the three co-aligned XMM-Newton X-ray telescopes has an
unimpeded light path to the primary focus, where the European Photon
Imaging Camera (EPIC) pn camera (Str\"{u}der \etal\ 2001) is
positioned.  The two other telescopes have Reflection Grating
Assemblies (RGAs) in their light paths, diffracting part of the
incoming radiation onto their secondary foci (where the Reflection
Grating Spectrometers (RGS; den Herder \etal\ 2001) are situated),
leaving the remainder to travel straight through to the primary foci,
where the two EPIC-MOS cameras (Turner \etal\ 2001) are positioned.

\begin{figure*}
\begin{center}
\includegraphics[width=6.5cm,angle=270]{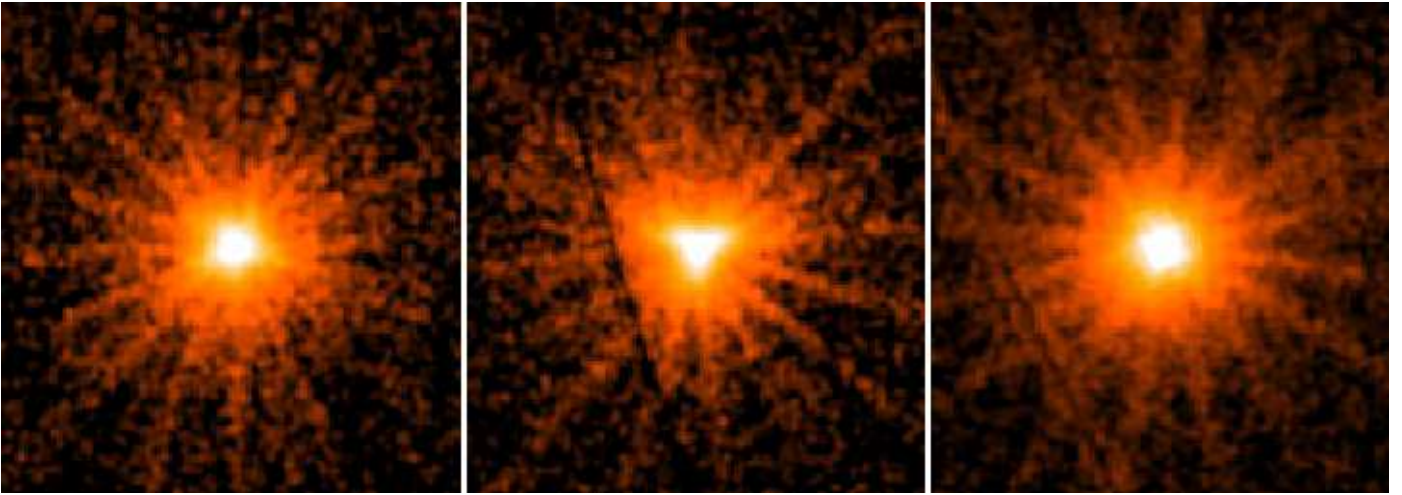}
\end{center}
\caption{PSFs of the 3 epic detectors (MOS1 left, MOS2 centre, pn
  right) for the same non-piled-up source from the same observation
  (2XMM~J130022.1+282402, ObsID 0204040101, revolution 823). The
  images are 0.2$-$10\,keV, are of binning 1.1\arcs$\times$1.1\arcs,
  and are very lightly smoothed, to accentuate the features. }
\label{3PSFs}
\end{figure*}

A critical parameter determining the quality of an X-ray mirror module
is its ability to focus photons, i.e. its Point Spread Function (PSF).
Each of the three Wolter type {\rm I} X-ray telescopes on board
XMM-Newton has its own PSF. As an example, Fig.\ref{3PSFs} shows the
in orbit on-axis PSFs of the MOS1, MOS2 and pn X-ray telescopes,
registered on the same on-axis non-piled-up source (specifically of
2XMM~J130022.1+282402 from ObsID 0204040101 from orbital revolution
823). This figure shows the shape of the PSFs, with the characteristic
radial spokes. Also note the coarser larger-scale shapes, in
particular the strong triangular form of the MOS2 PSF, and the weaker
pentagonal form of the MOS1 PSF. Note also that these coarse shapes
are quite different in the 3 EPIC cameras.

\begin{figure*}
\begin{center}
\includegraphics[width=8.85cm]{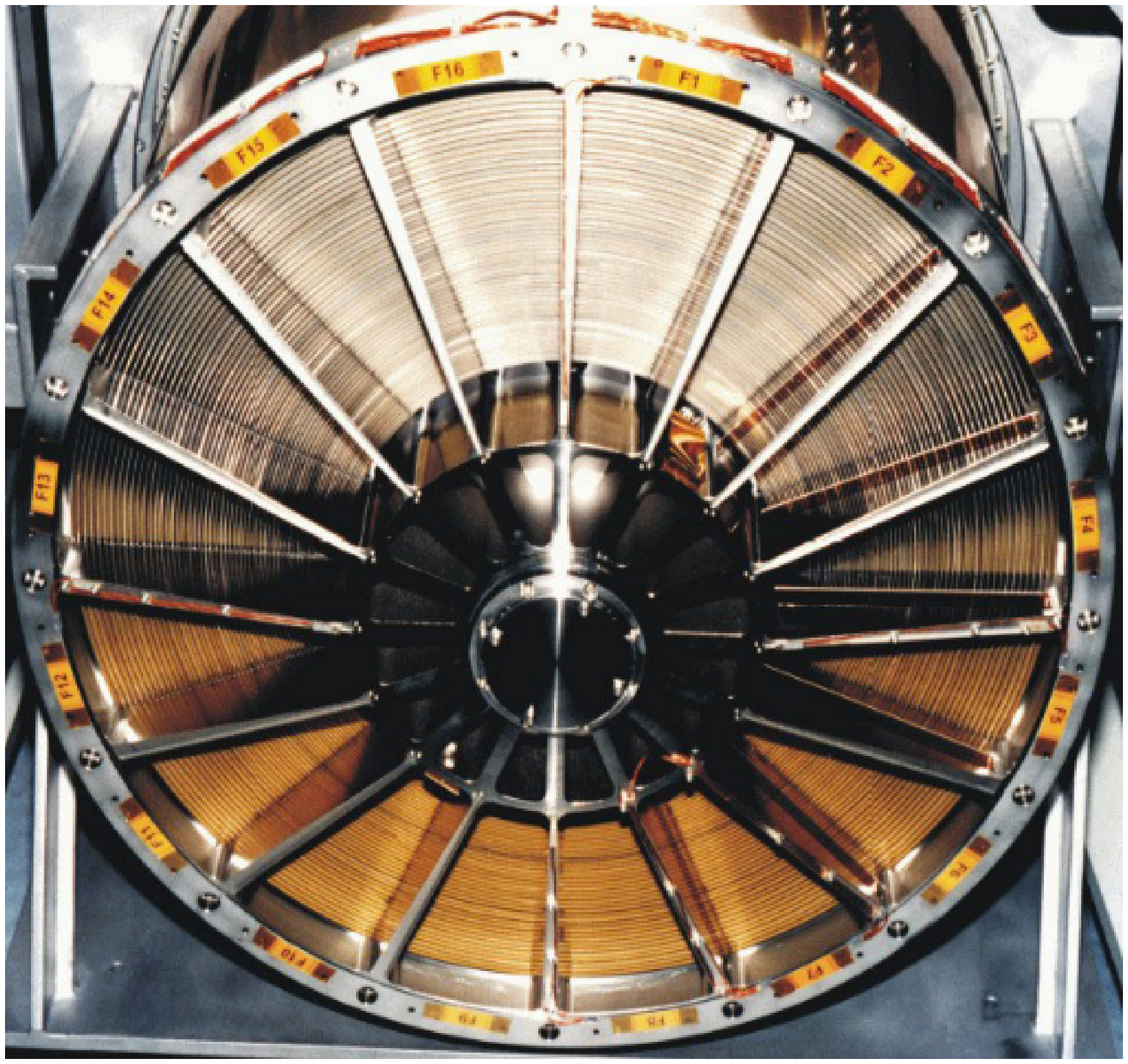}
\includegraphics[width=9.00cm]{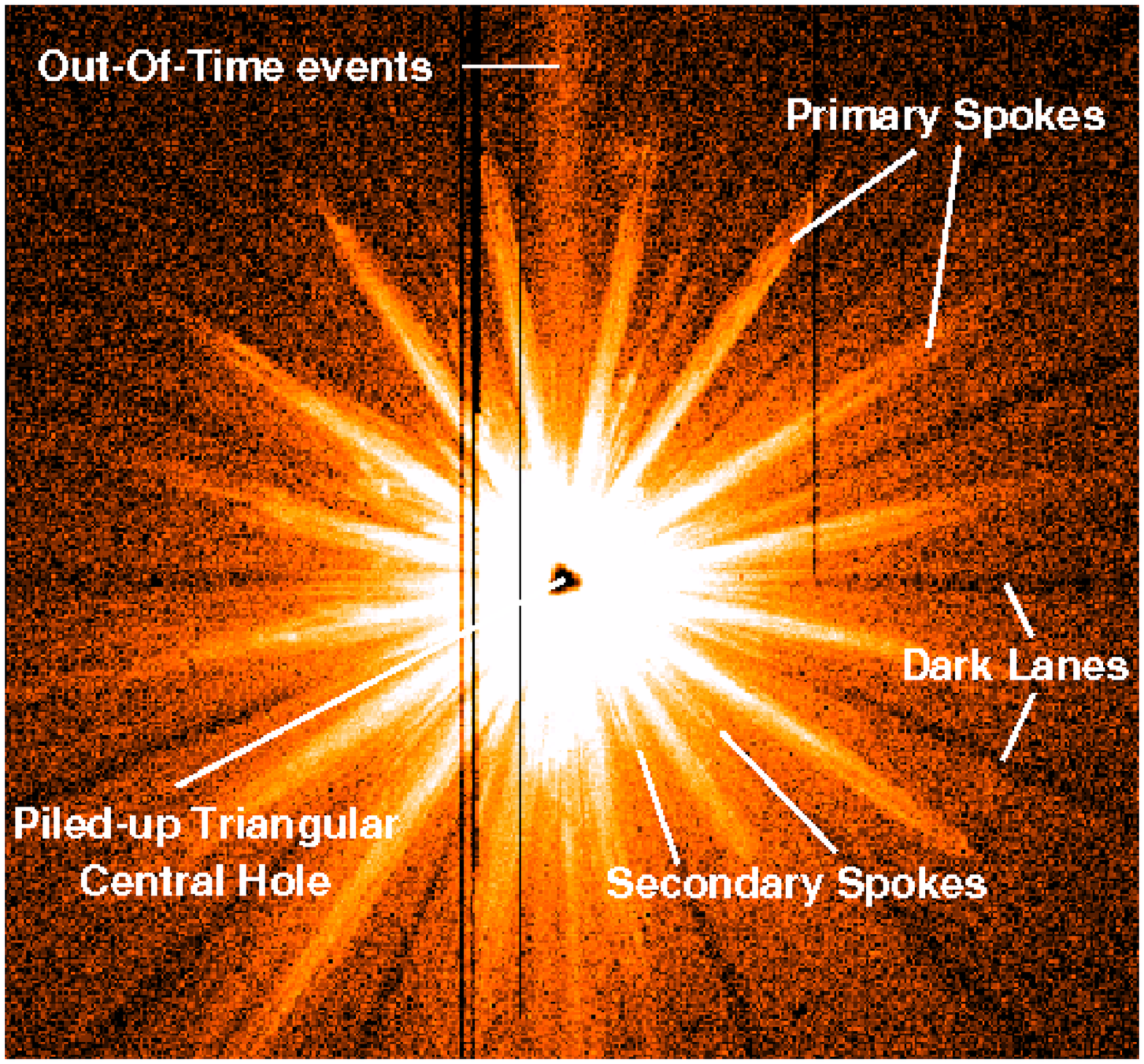}
\end{center}
\caption{(Left) A front-end view of one of the EPIC mirror modules
  containing the 58 co-axial mirrors shells, and the spider support
  structure used to hold the shells. (Right) The MOS2 PSF of the
  severely piled-up source GX~339-4 (ObsID 0204730301, revolution
  783), showing the various PSF and other features (see text) $-$ bad
  columns on the CCD are also visible as dark vertical lines.}
\label{mirror}
\end{figure*}

Any measurements of the PSF by the EPIC cameras may depend on the
instrument readout mode, through combinations of out-of-time (OOT)
event smearing and/or pile-up. The PSF can be severely affected by
pile-up effects when the count rate exceeds a few counts per
frame. Depending on the selection of event types in the EPIC event
analysis process, a hole can even appear in the core of the PSF due to
photon events undergoing severe pattern migration into unrecognized
and rejected patterns, or yielding a reconstructed energy above the
on-board high-energy rejection threshold. This is seen in
Fig.\ref{mirror} (right), where the PSF of a severely piled-up MOS2
on-axis source is shown (specifically of GX 339-4, ObsID 0204730301,
revolution 783).

Much more outer detail can be seen in the piled-up PSF (see
Fig.\ref{mirror}, right). Note for instance how the piled-up central
hole follows the general MOS2 triangular structure. The 16 primary
radial spokes are caused by the spider structure (see
Fig.\ref{mirror}, left) supporting the mirror shells (the primary
spokes in the image actually lie {\it between} the spider `legs',
i.e. were the spider absent, the PSF would be as bright as the primary
spokes around the whole azimuth). Note that the spider support
structure at the paraboloid front aperture is the only support
structure for the entire module; there is no equivalent counterpart
behind the hyperboloid rear aperture. Looking further, note the 16
lower-intensity secondary spokes lying between the primary
spokes. These are thought to be due to low-level scattering from the
sides of the spider legs. The dark lanes visible at larger radii from
the secondary spokes are due to the electron deflector which is
mounted after the rear aperture of the mirrors and whose legs are
aligned with those of the front-end spider. The coarser (triangular
and pentagonal) image structures are believed due to deformations in
the mirror shells, certain sets of mirror shells, believed to be the
outer shells (de Chambure \etal\ 1999), not being perfectly circular
in the 3 EPIC mirror modules. Finally, though not formally due to the
mirror system, note the diffuse streaks of OOTs to the top and bottom
of the piled-up PSF. These various structures to the EPIC PSF are
discussed throughout this paper.

\subsection{PSF descriptions within the XMM-Newton SAS}
\label{sec_psfdesc}

Historically there have been a number of descriptions of the EPIC PSF
that have been used as part of the XMM-Newton Scientific Analysis
System (SAS)\footnote{http://xmm.esa.int/sas/}. These include a number of two-dimensional (2-D) and
one-dimensional (1-D) descriptions. The only previous 2-D description
is the `Medium' mode PSF $-$ a set of simulated images in a matrix of
6 different off-axis angles and 11 different energies. This set of 66
images is identical for each EPIC instrument, and there is no
azimuthal variation included. This PSF description is the one that has
been used for source-searching throughout the XMM-Newton mission, and
is referred to as the `default' PSF in this paper.

1-D descriptions include (i) an `Extended' mode, incorporating a King
profile (with core radius and slope), as a function of EPIC
instrument, energy and off-axis angle, (ii) a `High' mode,
incorporating a 3-Gaussian parametrization of the `Medium' mode
images, and (iii) a `Low' mode, an early (and now unused)
single-Gaussian approximation of the PSF. The 1-D PSFs are often used
within the SAS for spectral work. These PSFs can all be found within the
particular XMM-Newton PSF current calibration files (CCFs). These are
named as XRT$i$\_XPSF\_$nnnn$.CCF, where $i$ refers to the instrument
(1=MOS1, 2=MOS2, 3=pn), and $nnnn$ is the issue number of the file.

That the default (Medium) PSF is limited, in that it is the same for
each EPIC instrument, and that no variation with source azimuth is
included, is a drawback. This is the PSF that has been used for
source-searching throughout the XMM-Newton mission, and it is likely
that a number of these limitations may be responsible for many of the
problems seen in the source-searching results and in constructing the
XMM-Newton catalogues (e.g. 2XMM; Watson \etal\ 2008), the major
problem being that a number of spurious sources are detected by the
standard source-detection pipeline, especially close to bright point
sources (note that post-pipeline visual screening does take place in
the production of the XMM-Newton catalogues, and sources that lie in
regions where spurious sources are considered likely to occur, are
flagged). There is therefore the need for a full 2-D energy- and
off-axis angle-dependent PSF model that incorporates all aspects of
the PSF structures, that is instrument-specific, and which accounts
correctly for a source's azimuthal position. This paper describes this
new 2-D PSF, and the structure of the paper is as follows: Sect.~2
describes the construction of the PSF $-$ the data analysis and the
modelling. Sect.~3 describes how the resultant new PSF model appears
in the SAS. In Sect.~4 we discuss our testing of the PSF, and
the major findings, and in Sect.~5 we present our conclusions.

\section{Constructing the 2-D PSF: Data Analysis and Modelling}

A standard method to obtain a fully 2-D characterization of the PSF as
a function of energy and off-axis angle across the entire
field-of-view (FOV) of each of the three EPIC instruments would be to
obtain a large number of images of appropriate point sources from the
XMM-Newton archive, and stack these together in
instrument/energy/off-axis angle groupings so that their shapes could
be analysed and modelled.

It quickly became apparent however that such an approach was
problematic. For a specific off-axis angle, the general
tangentially-stretched profile of the PSF that is naturally due to the
implementation of the Wolter {\rm I} optic rotates naturally around
the detector with source angle on the detector. The PSF details that
are to do with the support structure and the mirror deformations
however - i.e. the spokes and the triangles/pentagons) {\it do not}
rotate with source angle. These $-$ referred to collectively as the
support structure features $-$ remain fixed in detector angle as the
source rotates around the detector. An immediate upshot of this is
that the situation is very complex, with every single position on each
of the three EPIC detectors having a single unique PSF. Even for such
a high-throughput and long-term mission as XMM-Newton, nowhere near
enough good quality data exists to perform this full stacking analysis
at every detector position.

Consequently the new 2-D PSF system, both the modelling described here
and the incorporation of the PSF reconstruction into the SAS, has had
to be re-designed in the following way: The stackings in the
instrument/energy/off-axis angle groupings were used to construct
general elliptical `envelopes' (due to the optic implementation) for
that particular instrument, energy and off-axis angle, the support
structure features (the spokes, triangles etc.) having been averaged
and smeared out via the stacking procedure. 2-D spatial models were
then fit to these stacked envelopes to obtain general envelope spatial
parameters. Then, for a particular source at a particular known
azimuthal angle (whether on the detector, or on the sky), the
particular azimuthal support structures required (the spokes,
triangles etc.)  were folded into the appropriate elliptical envelope
to create the final 2-D PSF appropriate for that particular
instrument, energy, off-axis angle, and importantly, azimuthal
angle. This is described in more detail in the following sections.

\subsection{The Elliptical Envelope}

A major driver for this new 2-D description of the PSF was that it
should cover the entire FOV, i.e. both on-axis and
fully off-axis. It was necessary therefore to use data from those
observing modes that cover the full FOV - i.e. the `full-frame'
modes. Though smaller window modes can be useful in deriving some PSF
parameters (see later), these modes only exist on-axis, and it was
decided for the off-axis considerations, and for the sake of
consistency and uniformity, to select all the data solely from the
full-frame modes.

Sources were selected from the 2XMM catalogue (Watson \etal\ 2008) on
the basis of them:

\begin{itemize}

\item Being identified within 2XMM as point-like (i.e. no
  extension).

\item Having large numbers of (0.2$-$12\,keV) counts ($>$5000 for MOS1
  or MOS2 and $>$15000 for pn).

\item Having a countrate below the appropriate pile-up
  limits\footnote{XMM-Newton Users Handbook :
    http://xmm.esac.esa.int/external/xmm\_user\_support/documentation/uhb/}
  (0.70 ct/s [MOS full-frame], 6 ct/s [pn full-frame], 2 ct/s [pn
    extended full-frame]). 

\item Covering the full range in off-axis angle. (Note that the
  off-axis angle is just the `radial' angle relative to the optical
  axis, and any variation caused by mirror deformations (as opposed to
  support structure) with detector azimuth angle will be considered
  with the support structure features.)

\end{itemize} 

The appropriate raw data - the Observation Data Files (ODFs) - for
each source were identified and obtained, and the standard SAS (v7.0)
procedures (`epchain' for pn, `emchain' for MOS) were run on these to
create the standard calibrated event lists. These were then filtered
for periods of high background (solar proton flares) via standard Good
Time Interval (GTI) files created via analysis of high-energy
off-source lightcurves. Files (and therefore candidate sources) where
a large amount or long durations of high background flaring was
observed were rejected from further analysis. Large-scale images
around each source were created, and further rejections were made in
cases of crowded fields or where chip gaps or bad CCD columns were
seen to lie too close to the target source. 

For the sources that passed through to this stage
($>$250~observations, mostly comprising useful exposures for all three
EPICs, and several containing multiple valid sources), high-resolution
images were constructed around each source position. These were
constructed to be similar to the `Medium' PSF images (i.e. of
1.1\arcs\ resolution), and aligned such that the 2XMM source position
lay at the centre of the image. These images were created for (where
appropriate) each of the three EPIC cameras and in several different
energy bands; 0.1-1.0\,keV, 1.0-2.0\,keV, 2.0-3.5\,keV, 3.5-5.0\,keV,
5.0-7.0\,keV, 7.0-9.0\,keV, 9.0-11.5\,keV, 11.5-15.0\,keV. All other
sources (all of which were very faint and well separated from the
target source) were excised from each image (using a circle of radius
35\arcs), the holes filled with the average value in the surrounding
(40\arcs$-$70\arcs) annulus. Each image was then rotated with respect
to the source's azimuthal angle on the sky (this angle is equivalent
to a combination of the position angle of the observation and the
source's azimuthal angle on the particular detector). A Gaussian was
fit to the rotated image to account for any tiny shifting in the
procedure (always $<<$1\arcs), and a final rebinning of the rotated
image about this fit centre was performed.

The final good images were collected together for each combination of
EPIC instrument (3), energy band (8, linear fiducial midpoints:
0.55\,keV, 1.5\,keV, 2.75\,keV, 4.25\,keV, 6.0\,keV, 8.0\,keV,
10.25\,keV, 13.25\,keV) and offaxis-angle band (6; 0-1.5\arcm,
1.5-4.5\arcm, 4.5-7.5\arcm, 7.5-10.5\arcm, 10.5-13.5\arcm, \&
$>$13.5\arcm, these assigned to the fiducial values 0\arcm, 3\arcm,
6\arcm, 9\arcm, 12\arcm, 15\arcm). These sets of images were then
stacked together, bringing them together to a common reference frame.

These stacked images were fit with a 2-D King profile using the {\tt beta2d}
model in {\sc CIAO-sherpa\footnote{http://cxc.harvard.edu/sherpa/}} :
$$
B(r)=\frac{A}{[ 1+(r/r_0)^2]^{\alpha}}
$$
where:
$$
r(x,y,\theta) = 
$$
$$
\sqrt{[(x \cos \theta + y \sin \theta)^2 ] + [(y \cos \theta - x \sin \theta)^2] / (1 - \epsilon)^2 } 
$$

and $r_0$ (core radius), $\alpha$ (index), $\epsilon$ (ellipticity) and
$\theta$ (angle of ellipticity) are the model parameters.

\begin{figure}
\begin{center}
\includegraphics[bb=107 56 505 739, width=8.8cm]{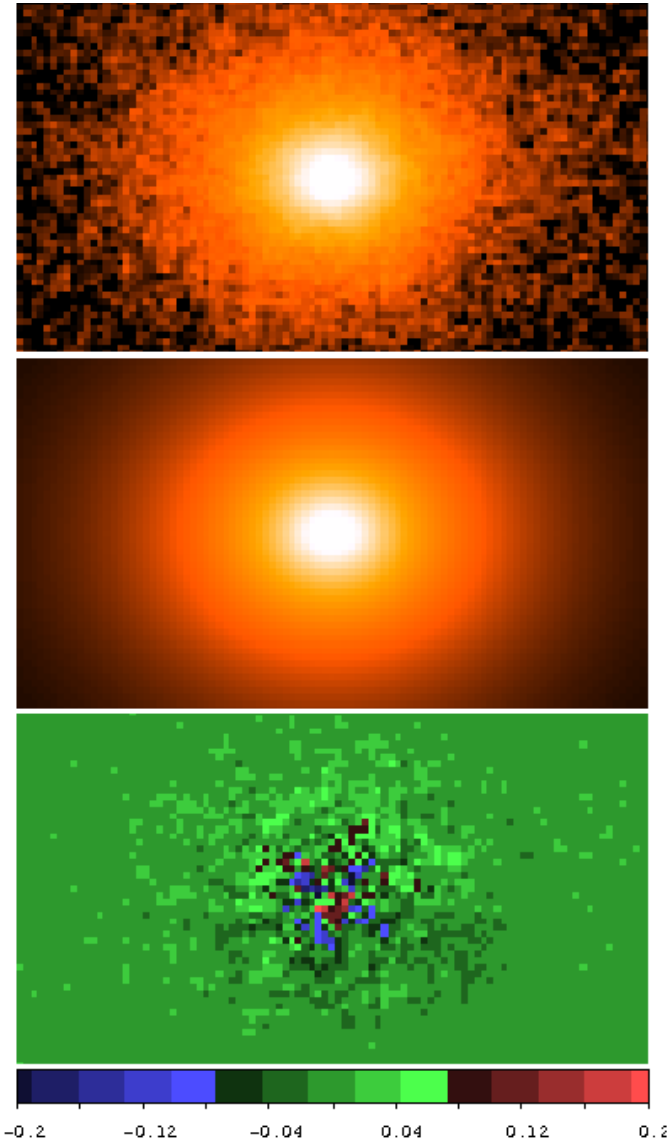}
\end{center}
\caption{Results from a typical run of the 2-D fitting (specifically
  for pn, 6\arcm\ off-axis angle, 1.5\,keV). The top and middle panels
  show the data and final model respectively, at the same scaling. The
  bottom panel and colourbar (running from -0.2 to 0.2) show the
  residuals (data$-$model, the central peak of the data and model
  panels being $\approx$5 in the colourbar units). }
\label{resid}
\end{figure}

To the parametrisation above, a further 2-D Gaussian core has been
added to model excess emission in the low- and medium-energy ($E \le
6$~keV) PSF at the very centre of the MOS cameras (no such core is
observed to be necessary for pn). The model includes the same
ellipticity term through the definition of the $r$ variable as above.
The 2-D Gaussian function used ({\em gaus2d} model in {\sc
  CIAO-Sherpa}) is :
$$
G(r) = A e^{-4 \ln(2) (r/{\rm FWHM})^2}
$$ where FWHM is the full width at half maximum. All these parameters
are contained within the PSF CCFs (issue numbers 0011 and
after). Finally, a flat background level, fixed at the mean value
beyond a 5\arcm\ radius was added to the model for each stacked image.
For all instruments and off-axis angles, the two highest energy
stacked images (at 10.25\,keV and 13.25\,keV) were seen to contain
only very sparse data, and it proved impossible to obtain stable 2-D
fits with sensible error bounds in many of these cases. Consequently,
the images in the two highest-energy bands have been combined together
for all instruments and off-axis angles, and the resultant spatial fit
parameters obtained have been applied to both energy bands in the
resultant calibration files (at 10.25\,keV and 13.25\,keV). Future
calibration, using more data, may be able to establish fit parameters
for both energy bands separately. An example showing the results from
a typical run of the 2-D fitting (specifically for pn,
6\arcm\ off-axis angle, 1.5\,keV) is shown in Fig\ref{resid}.

The actual parameter values, which are contained within the PSF CCFs
(described in more detail in Sec.\ref{sec_ccf}), have evolved, and
will continue to evolve, with iterations in the analysis and
modelling. It is possible however to discuss some general trends in
the current parameters (CCF issue number 0013). Dealing first with
$r_0$ (core radius) and $\alpha$ (index), note that there is some
degeneracy between these two parameters; increasing (or decreasing)
both parameters simultaneously can sometimes lead to only very small
changes in the overall profile. The two MOS PSFs behave similarly;
close to on-axis, $r_0$ is seen to gently decrease with increasing
energy, while $\alpha$ is seen to remain roughly constant with
energy. The profile therefore is seen to get narrower at higher
energies, as expected, as only a fraction of the mirror shells $-$ the
smaller, inner shells $-$ contribute to the higher-energy PSF. At
larger off-axis angles, the MOS $r_0$ and $\alpha$ values are seen to
decrease with increasing energy, but more steeply, starting from
larger low-energy values than is seen on-axis. The pn PSF behaves
slightly differently in that $r_0$ decreases gently with increasing
energy for all off-axis angles, but the average $r_0$ value increases
with off-axis angle (as expected, the PSF becoming wider
off-axis). The pn $\alpha$ is seen to remain roughly constant for
almost all energies and off-axis angles, hence, in conjunction with
the gently decreasing $r_0$ values with energy, the profile again gets
narrower at higher energies. The ellipticity behaves as one would
expect and very similarly for each instrument, getting larger with
increasing off-axis angle, from $\approx$0 on-axis to $\approx$0.6 far
off-axis. Also the ellipticity is seen (more prominently off-axis) to
gently increase with increasing energy in each instrument. Finally,
for the Gaussian component (which exists only for the MOS cameras at
$E \le 6$~keV), both the FWHM and the relative normalization for both
MOS PSFs are seen to fall from 0 to 6\,keV, for all off-axis angles
(i.e. to match there being no Gaussian at $>$6\,keV).

\subsection{The Support Structure features - The Spokes}

\begin{figure*}
\begin{center}
\includegraphics[bb=0 190 720 540,width=16cm]{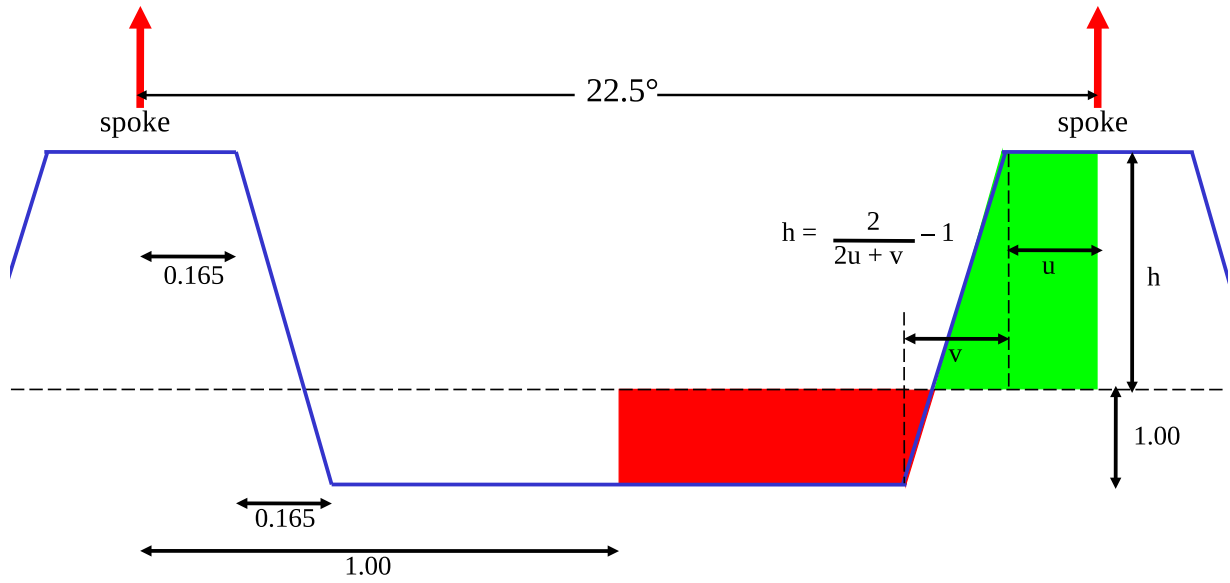}
\end{center}
\caption{Schematic of the flat-topped triangular function used to
  parameterize the PSF spokes. The blue spoking function is
  constructed such that the red and the green areas exactly cancel
  out. The dashed line shows the unspoked function. The lengths 1.00,
  0.165 (the half-width of the flat-top $u$) and 0.165 (the base to
  peak width $v$) are in units of half the inter-spoke distance, {\it
    i.e.} 11.25$^{\circ}$. The figure is not to scale.}
\label{flattop}
\end{figure*}

The radial spokes have been modelled with a flat-topped triangular
function (Fig.~\ref{flattop}), chosen after consultations with mirror
experts (B.\,Aschenbach, private communication; R.\,Willingale,
private communication). 16 equally-spaced spokes have been included in
the model. The distance between two consecutive peaks of the spoke
function is therefore 22.5$^{\circ}$. The shape of this function has
been tuned in such a way that it does not change the integral of the
radial profile ({\it i.e.}, the red and the green areas in
Fig.~\ref{flattop} exactly cancel out). This filtering function is
fixed in space with respect to the detector, such that any axis of the
DETX/Y (or RAWX/Y) coordinate system lies exactly between two spokes,
and is only applied to the elliptical envelope {\it after} the
envelope has been rotated according to the azimuthal position of the
source on the detector.

The current spoke dimensions have been obtained from measuring the
azimuthal angles of the spoke troughs, peaks and flat-tops of a number
of bright sources, including GX~339-4, using the full energy range
(0.2-12\,keV) and over a large radial range (20\arcs$-$100\arcs). The
current dimensions are such that both the half-width of the flat-top
($u$), and the width from base to peak ($v$) equals 0.165 times half
the width of the spoke-to-spoke distance (as indicated in
Fig.\ref{flattop}). It can be shown that the height $h$ of the spoke
above the unspoked dashed line of Fig.\ref{flattop} (if the spoke
drops below the line by a depth of 1.0) is related to the half-width
of the flat-top $u$ and the base to peak width $v$ via :
$$
h=\frac{2}{2u+v} - 1
$$

The strength (intensity) of the primary spokes is set such that the
spoked PSF is made up of (at most) 42\% of an image formed from the
flat-topped triangular function depicted in Fig.\ref{flattop}, and the
remainder is formed from an unspoked image. This is again based on
the spatial analysis of a number of bright sources. The radial
dependence of the intensity of the spoking is however not constant
(see below), and this intensity value (42\%) is the maximum value.

16 secondary spokes, observed in very bright sources (see
Fig.\ref{mirror}), and situated half-way between the 16 main primary
spokes, are modelled with the same spoke function, rotated by
11.25$^{\circ}$ with respect to the main spoke system, and with a
(maximum) intensity factor of 3.5\%.

The relative strength of the spokes is not constant with radius but
appears to vary. Owen \& Ballet (2011) were able to characterize the
radial dependence of the spokes of the MOS1, MOS2 and pn PSFs to large
radii (up to 2\arcm) using bright, heavily piled-up sources, and using
procedures developed to correct for the effects of pile-up. They found
that the radial dependence of the spokes of the two MOS cameras and
the pn camera appeared to be approximately the same. This is not
surprising as the spokes and their radial dependencies are all due to
the construction of the mirror modules, all of which were designed to
be identical. Furthermore the spiders were constructed out of Inconel,
a material chosen for its thermal expansion, which is close to that of
the electrolytic nickel of the mirrors (de Chambure \etal\ 1999), thus
keeping distortions to a minimum. This radial dependence has been
modelled to match the Owen \& Ballet (2011) on-spoke to off-spoke
ratio results, and will be introduced into the PSF CCFs (for issue
numbers 0014 and after) as follows: For each EPIC PSF, from the centre
out to 10\arcs, there is no spoking. From 10\arcs to 110\arcs\ the
spoke strength increases linearly from zero to the maximum
(42\%). From 110\arcs\ to 180\arcs\ the spoke strength decreases
linearly from its maximum to zero, and there is no spoking beyond this
radius.

\begin{figure}
\begin{center}
\includegraphics[bb=85 65 540 510, width=8.5cm]{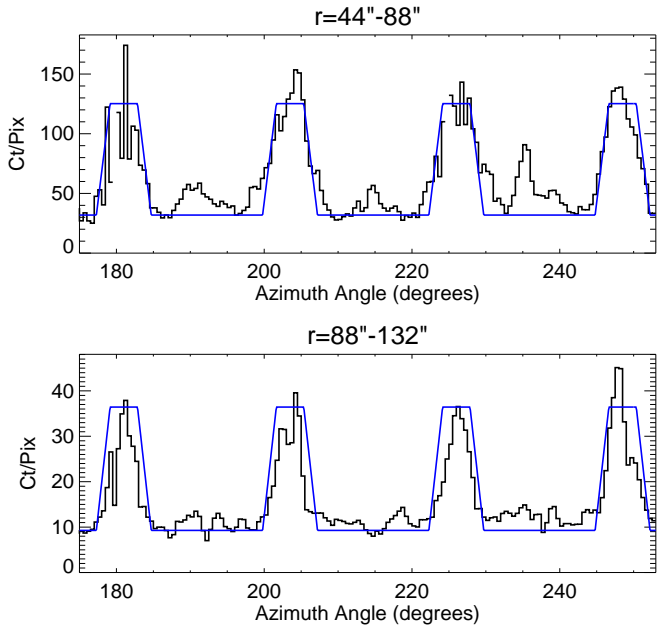}
\end{center}
\caption{A section of the azimuthal profile (black line) from GX~339-4
  (ObsID 0204730301) (MOS1, 0.2-12\,keV) in annular extractions of
  (top) 44\arcs$-$88\arcs and (bottom) 88\arcs$-$132\arcs. The current
  (CCF 0013) flat-topped triangular function for the primary spokes at
  maximum strength (i.e. at a radius of 110\arcs) is shown in blue
  (see text).}
\label{azimcomp}
\end{figure}

Looking beyond the present model, when we compare in detail the
azimuthal profile of good quality point source data with the current
(CCF 0013) model (see Fig.\,\ref{azimcomp}), there are a number of
points of interest: Though there is significant statistical scatter,
there may also be some true spoke-to-spoke variation. Further, it
appears that there may be some variation of the spoke model width
parameters $u$ and $v$ with radius $-$ the spokes appearing to be
narrower at larger radius. Interestingly it appears that there is a
significant difference in the radial dependence of the secondary spoke
strength, compared to that of the primary spokes $-$ the secondary
spokes falling off quicker (there also appears to be significant
secondary spoke-to-spoke variation). None of these effects (nor their
energy-dependencies) are yet contained within the current PSF model,
but they impinge upon e.g. the on-spoke to off-spoke Owen \& Ballet
(2011) results, and on their interpretation and modelling $-$ note
that the on-spoke to off-spoke ratio is lower at smaller radius
(44\arcs$-$88\arcs) than at larger radius (88\arcs$-$132\arcs), partly
because the secondary spokes are relatively stronger at smaller radius
than at larger radius. It is hoped that these issues can be revisited
in future improvements in the modelling of the EPIC PSFs.

\subsection{The Support Structure features - The Large-scale azimuthal modulation}

The apparent triangular shape of the MOS2 PSF can be modelled by a
spatial low frequency overall modulation of the PSF shape. A
pentagonal modulation is also present at a lower level in the MOS1 and
also at a very low level in the pn PSF. These shapes are thought to be
due to distortions and irregularities in certain sets of mirror shells
within each of the telescopes - e.g. the outer shells not being
perfectly circular (de Chambure \etal\ 1999). It is not surprising
therefore that these deformities are different in the three EPIC PSFs.

This modulation $M(\phi)$ is modelled as a function of the azimuthal
angle $\phi$, with a multi-peaked cosine function in the PSFs of the
cameras:
$$
M(\phi) = A \cos (N\phi - \phi_0 ) 
$$ where $A$ is the amplitude of the modulation, $N$ is the number of
peaks in the modulation in a full 360$^{\circ}$ revolution, and
$\phi_0$ is the azimuthal offset of the cosine function. The current
parameters of this function are shown in Table~\ref{grossaz}.

\begin{table}
\caption{Current parameters of the large-scale azimuthal multi-peak
  modulation. The modulation $M(\phi)$ is modelled as a cosine
  function of the azimuthal angle $\phi$, with $A$ as the amplitude of
  the modulation, $N$ being the number of peaks in the modulation in a
  full 360$^{\circ}$ revolution, and $\phi_0$ the azimuthal offset of the cosine
  function. $\phi$ and $\phi_0$ run clockwise from north for a source
  on the sky, for an observation Position Angle, PA~=~0 (see text).  }
\label{grossaz}
\begin{center}
\begin{tabular}{lccc} \hline \hline
Cameras & $A$ & $N$ & $\phi_0$ \\ \hline
MOS1 & 13\% & 5 & 62$^{\circ}$ \\
MOS2 & 45\% & 3 & 50$^{\circ}$ \\
\hline \hline
\end{tabular}
\end{center}
\end{table}

The radial dependence of these large-scale azimuthal modulations is
also seen not to be constant with radius, and the maximum amplitude
$A$ is listed in Table\ref{grossaz}. The dependence is only roughly
similar to the radial dependence of the spokes, and there are
variations between the EPIC detectors - the pn pentagon for example
only appears to be visible over quite a narrow range in radius. This
radial dependence of the large-scale azimuthal modulations (and the
entire pn large-scale azimuthal modulation) has not been fully
calibrated yet, but will appear in a future issue of the PSF CCFs.

As a final step in the construction of the 2-D PSF, a
radially-dependent very light smoothing is applied. A smoothed image
of the PSF is constructed using a flat boxcar of halfwidth
1.65\arcs. The final PSF is then a sum of $f$ times this smoothed
image plus $(1-f)$ times the original unsmoothed image, where $f$
varies linearly from 0 at the source center to unity at
8.8\arcs. Though this does not currently directly match the more
elliptical nature of the PSF at larger off-axis angle, this will be
investigated further, and the smoothing effect is very small, has been
applied merely to de-pixelate the spoke structures at large radius,
and does not have any significant large-scale effects (e.g. on radial
profiles).

\subsection{Additional features}

The new 2-D PSF system has been designed to be modular, and it is
quite straightforward to add in further complexities. There are a
number of additional features to the EPIC PSFs that will be added, it
is intended, to future updates of the system (some may just require
updates to the CCFs, others may require software and infrastructure
changes). The radial dependencies of the large-scale azimuthal
modulations, as discussed above, is one. It is very probable, as the
large-scale azimuthal modulations are due to deformations only in
certain mirror shells, that there are energy-dependencies to this
effect, and these will also need to be incorporated. A further task is
to include the effects of the dark lanes visible e.g. in
Fig.\ref{mirror}, due to the electron deflector, mounted after the
rear aperture of the mirrors and whose legs align with those of the
front-end spider. It should be possible, once calibrated, to insert
this effect into the system, via a similar azimuthal filtering
technique as for the spokes. Features due to OOTs have not been
considered to be part of the PSF. Though instrument-dependent, they
are also mode-dependent, and in terms of the source-detection
software, they are modelled via an alternative route (at least for the
pn, where the OOT signal is much stronger). The OOT features here
have, in a similar manner to the azimuthal structures (the spokes,
triangles etc.), been diluted and smeared out via the stacking
procedure prior to the envelope fitting.  Furthermore there is set
within the PSF system the possibility to allow for any azimuthal
variations in the envelope 2-D King and Gaussian parameters. Currently
the CCFs are set such that these parameters do not vary with source
azimuth, but future calibration may require changes, e.g. to account
for variations due to the obscuring RGAs in the EPIC-MOS cameras.

\section{Results: The 2-D PSF in the XMM-Newton SAS}

The new 2-D PSF is contained within all XMM-Newton SAS PSF Current
Calibration Files (CCFs) of the form XRT$i$\_XPSF\_$nnnn$.CCF, where
$nnnn$, the issue number of the file, is 0011 or higher. The relevant
parameters are contained within the {\tt ELLBETA\_PARAMS} extension.

A number of elements are used to describe the new 2-D PSF in these
CCFs, as summarized here: 

\begin{itemize}

\item a (currently) azimuthal-angle-independent parameterization of the
  elliptical PSF envelope as a function of instrument, energy, and
  off-axis angle.  An additional Gaussian `core' is added to the PSF
  envelope of the MOS cameras for low and medium energies ($E \le
  6$~keV) only.

\item an azimuthal filter of the elliptical envelope, which describes
  the effect of the spokes created by the spider supporting the 58
  co-axial mirrors of each telescope. Primary and secondary spokes are
  included. The radial dependency of the strength of these spokes is
  to be included in version 0013 of the CCFs. The parameters of this
  filtering are currently hard-coded into the CAL software, but will
  appear as FITS keywords in version 0013 of the CCFs. The filtering
  to account for the dark lanes will appear in future CCF releases.

\item a further azimuthal dependency of the overall PSF envelope,
  which is responsible for the apparently triangular shape of the MOS2
  PSF. Lower level similar effects are present in the PSFs of MOS1 and
  pn also. Currently, this correction is applied to the PSFs of MOS1
  and MOS2 only. The parameters of this correction are currently
  hard-coded into the CAL package, but will appear as FITS keywords in
  version 0013 of the CCFs. The pn correction and the radial
  dependencies for all three will be included in future releases of
  the CCFs.

\end{itemize}

\begin{figure*}
\begin{center}
\includegraphics[width=8.75cm,angle=270]{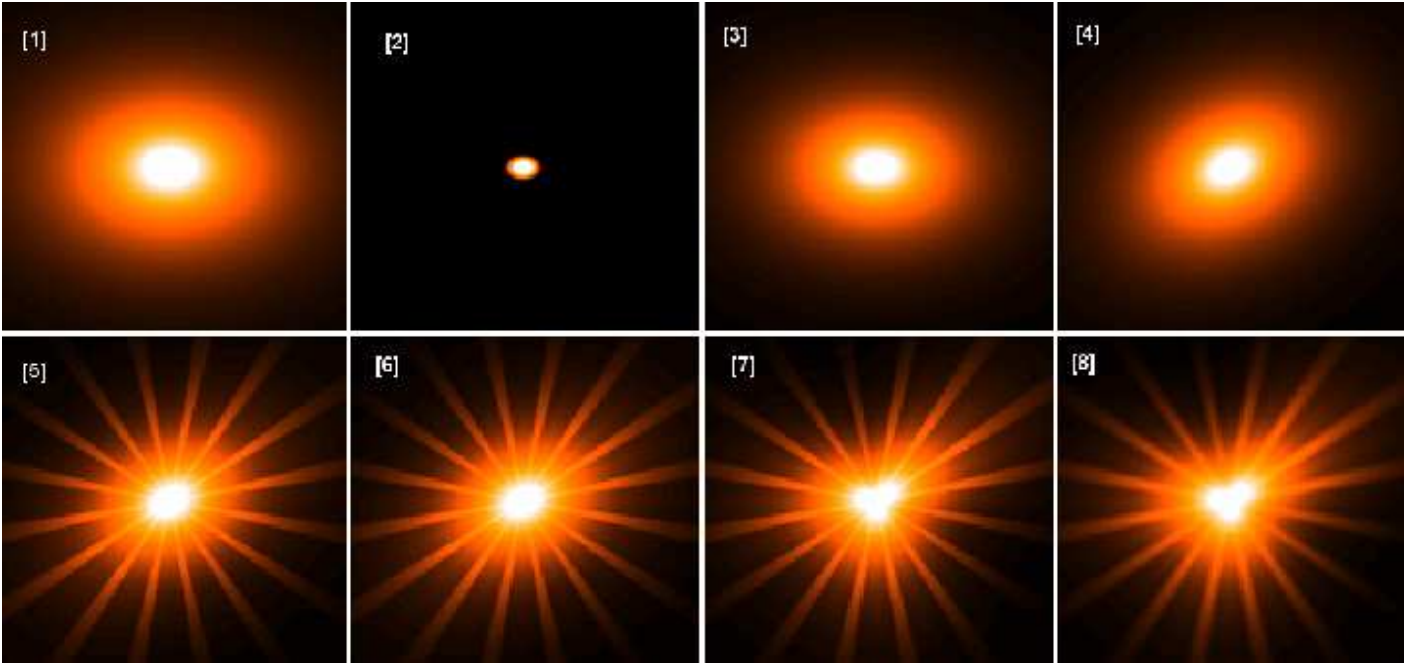}
\end{center}
\caption{The eight main steps in the formation of the full 2-D PSF for
  a source in a given instrument, of a given energy and at a given
  off-axis and azimuthal angle: The King ({\tt beta2d}) component [1]
  is constructed, then the Gaussian ({\tt gaus2d}) core [2] is
  constructed, and these are added [3] in the correct ratio (the CCF
  parameters in steps 1-3 are all functions of instrument, energy and
  off-axis angle). Then this is rotated [4] according to the azimuthal
  position of the source on the detector, and only then are the
  radially-dependent primary [5] and secondary [6] spoke structures
  azimuthally filtered in, using a flat-topped triangular
  function. Finally, the large-scale azimuthal modulation (a function
  of EPIC instrument) is filtered in [7], and the very light
  radially-dependent smoothing applied [8]. The example shown is for
  MOS2, at an energy of 1.5\,keV, an off-axis angle of 9\arcm, and a
  source azimuthal position of 30$^{\circ}$. }
\label{8steps}
\end{figure*}

\begin{figure}
\begin{center}
\includegraphics[width=8.75cm]{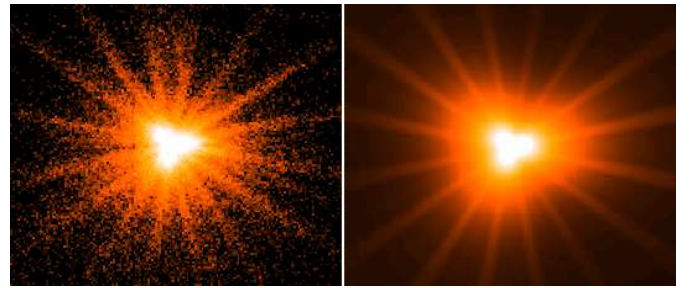}
\end{center}
\caption{A very bright, slightly piled-up, $\approx4$\arcm\ off-axis
    angle MOS2 point source and the equivalent PSF model at a similar
    off-axis and the appropriate source azimuthal position.}
\label{psfcomp}
\end{figure}

A scheme of the different physical ingredients of the 2-D PSF
parameterization and how they are combined to produce the final PSF is
shown in Fig.~\ref{8steps}. For a comparison with an actual EPIC
off-axis source, Fig.~\ref{psfcomp} shows a very bright,
$\approx4$\arcm\ off-axis angle MOS2 point source and the equivalent
PSF model at a similar off-axis and the appropriate source azimuthal
position.

Once the 2-D PSF was incorporated into both the CCF system and the
SAS, a next step was to compare it with the default PSF in terms of
how well the SAS-constructed PSFs model the true source structures. To
this end, full SAS calibrated event list creation, image formation and
detection chain analysis (as per the standard pipeline) was performed
on a large number of fields containing bright, non-piled-up point
sources, detected by EPIC, and (where available and where observing
mode constraints allowed) in all three EPIC cameras. The detection
chain analysis was performed once using the default (MEDIUM) PSF and
once using the 2-D (ELLBETA) PSF. Next, taking great care to co-align
each source image and each detection-chain-produced model PSF image,
then for each source, both the source and the model counts within each
of 30 radial bins (each 4\arcs\ wide) and 32 azimuthal bins (each
11.25$^{\circ}$ wide, arranged to be alternately fully on-spoke, then
fully off-spoke) were computed. The co-alignment ensured that each
azimuthal bin corresponded to the same azimuth on the PSF. This
procedure was performed for 80 bright, non-piled up point sources, and
the values were stacked together, in several energy bands and off-axis
angle groupings. All this analysis was performed using both the 2-D
PSF and the default PSF.

\begin{figure*}
\begin{center}
\includegraphics[width=9cm,angle=270]{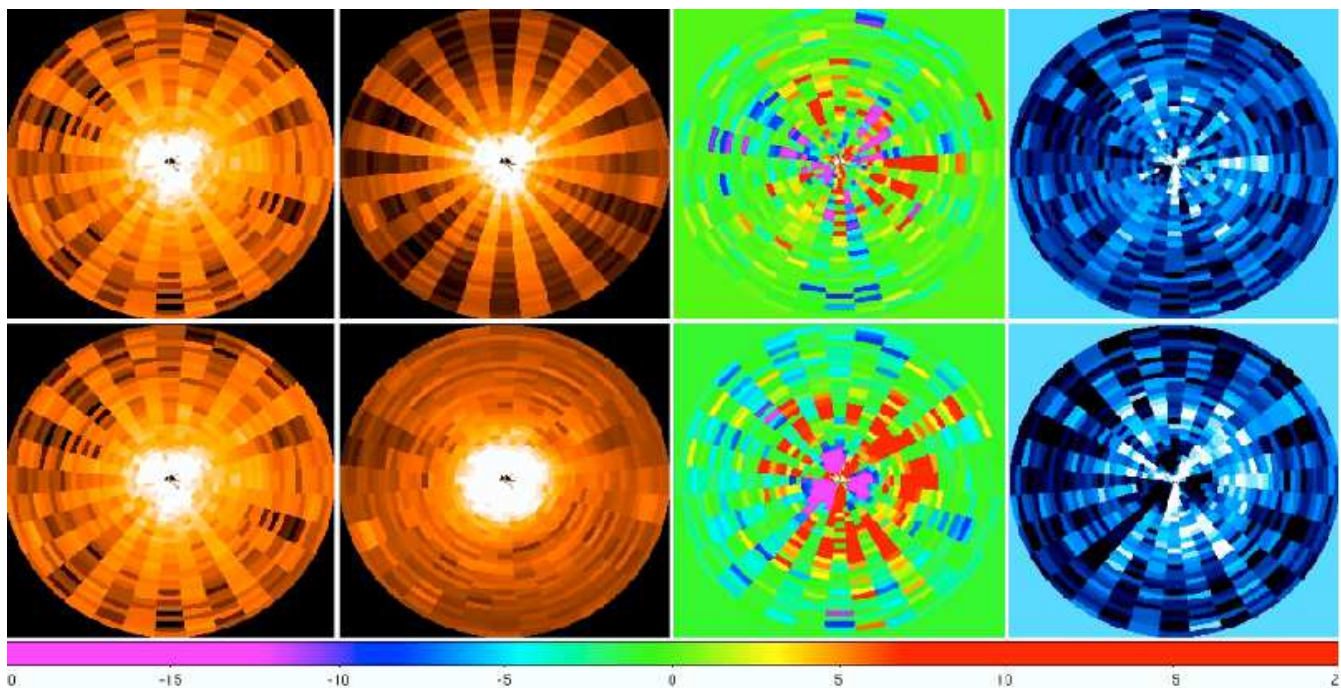}
\end{center}
\caption{Comparing the SAS-produced 2-D PSFs and default PSFs with
  true source data. Left to right: (1) the stacked counts in the data,
  (2) the stacked counts in the model, (3) a $\chi^{2}$-like
  statistic, (4) $r$, a measure of whether the model consistently
  (white) under- or (black) overestimates the data (see text for all
  details), for (top) the 2-D PSF and (bottom) the default PSF. The
  colourbar corresponds to the $\chi^{2}$ plot. The 30 radial bins are
  each 4\arcs\ wide, and the 32 azimuthal bins are each
  11.25$^{\circ}$ wide, and arranged to be alternately fully on-spoke,
  then fully off-spoke. The example shown is for MOS2, on-axis,
  0.5$-$1\,keV. }
\label{dmcr}
\end{figure*}

Fig.\ref{dmcr} shows the results of this analysis, for the example of
MOS2, on-axis, 0.5$-$1\,keV, arranged in the circular grid used of 30
radial bins combined with 32 azimuthal bins. The top row corresponds
to the 2-D PSF, and shows (left to right): (1) the stacked counts in
the data, (2) the stacked counts in the model, (3) a $\chi^{2}$-like
statistic, calculated from the stacked data counts and stacked model
counts as :
$$
\chi^{2} = \frac{\pm (\mbox{stacked data} - \mbox{stacked model})^{2}} {\mbox{stacked model}},  
$$ and (4) $r$, a measure of whether the model consistently under- or
overestimates the data, and calculated as the sum of the individual
data$-$model `signs' via :
$$
r = \sum \frac{\rm{data-model}}{|\rm{data-model}|}, 
$$ which ranges between +1 (white, always underestimates) and -1
(black, always overestimates). The bottom row of Fig.\ref{dmcr} shows
the equivalent plots for the default PSF (and hence the two stacked
data plots are identical). The colourbar (-20 to +20) corresponds to
the $\chi^{2}$ plot.

A number of conclusions can be drawn from this figure. Similar results
are seen for MOS1 and pn, though some (e.g. to do with the large-scale
azimuthal filtering) are harder to see. Very obvious is that the
default PSF, due to its limitations discussed in
Sec.\ref{sec_psfdesc}, performs very poorly with regard to the spokes;
the on-spoke regions are very underestimated by the default model, and
the off-spoke regions are overestimated. Secondly, the default PSF
performs extremely badly in modelling the large-scale azimuthal
features $-$ here, the MOS2 triangle (note the very large triple-peaked
discrepancies in the $\chi^{2}$ and $r$ plots for the default
PSF). The 2-D PSF however, performs very much better. Much less
variation in $\chi^{2}$ and $r$ is seen on- and off-spoke. The plot
does suggest though that the 2-D PSF relative spoke strength may be
too strong. The most up-to-date SAS and the latest PSF CCF files that
are available for this analysis (version 0012) however, do not yet
include the radial dependency of the spoke strength, and the future
inclusion of this will act to decrease this effect, and improve
further the 2-D PSF modelling. The 2-D PSF modelling of the MOS2
triangle is also very much improved over the default PSF. Further
refinements to this component include the modelling of both its
radial-dependence and its energy-dependence.

\subsection{Description of the PSF CCF parameters}
\label{sec_ccf}

The FITS extension {\tt ELLBETA\_PARAMS} in the PSF CCF
XRT$i$\_XPSF\_0011.CCF (and later) files contains the parameters
describing the elliptical envelope and the Gaussian core.  This
extension contains four columns:

\begin{itemize}

\item ENERGY: the energy (in eV) to which the parameters
refer

\item THETA: the off-axis angle (in radians)
to which the parameters refer

\item PHI: the azimuthal angle (in radians)
to which the parameters refer

\item PARAMS: an array, containing the parameters of the 2-D King
  ({\tt beta2d}) plus Gaussian ({\tt gaus2d}) function:

\begin{itemize}

\item {\tt 1}: the King core radius ($r_0$), in arcseconds

\item {\tt 2}: the King power-law slope ($\alpha$)

\item {\tt 3}: the ellipticity ($\epsilon$) (of both the King and the Gaussian components)

\item {\tt 4}: the Gaussian Full Width Half Maximum (FWHM), in arcseconds

\item {\tt 5}: the normalisation ratio of the Gaussian peak to the King peak ($N$)

\end{itemize}

\end{itemize}

\section{Discussion: Testing of the PSF }

Though this paper is primarily concerned with a description of the new
2-D PSF and its construction etc., it is very useful here to discuss
some of the testing, especially as regards the major reason why the
new 2-D PSF was originally desirable - i.e. to see if it could reduce
the number of spurious sources detected by the standard
source-detection pipeline, where the relatively poor description of
the PSF can lead to large residuals in the constructed
data-minus-model images, leading in turn to the detection of spurious
sources. Also of major interest are the astrometry issues that have
arisen as a consequence of these first tests. These issues are
discussed in the following sections.

\subsection{Source-searching and spurious sources}

A comparison of the performance of the new 2-D PSF (`Ellbeta') with
that of the current existing default PSF (`Medium') has been
performed, the primary objective being to establish whether the 2-D
PSF better suppresses the detection of spurious sources. The analysis
first focussed on a set of 108 `problem' observations containing
examples of various types of situations where spurious sources had
previously been found to be present. These were mainly chosen to be
around non-piled-up bright sources, but there were also several fields
chosen involving bright piled-up sources, off-axis sources, extended
sources, out-of-time (OOT) events from piled-up sources and single
reflection arcs. A separate `random' set of 83 observations was also
processed as a reference, as was a `clean' subset of 40 fields, free
of any `problems', extracted from the `random' subset.

Standard calibrated event file creation, and background cleaning was
performed on the datasets, and standard images were extracted from the
data in the 5 usual energy bands used for EPIC source-detection in the
2XMM catalogue (Watson \etal\ 2008). The main source-detection task
emldetect (v5.15) was then used, once using the default PSF, and once
using the 2-D PSF. Importantly, exactly the same input files
(specifically the source images, background images, exposure maps,
detector masks and eboxdetect input lists) were used in the default
and 2-D runs, ensuring that no changes were introduced due to
e.g. spatial randomization of the events, and that any changes seen
would be due solely to the PSF usage.

The numbers of emldetect detections obtained using the default and the
2-D PSFs for the `problem' fields (split into the various problem
groups), and for the `random' fields and the `clean' fields are shown
in Table~\ref{sourcesearch}.

\begin{table}
\caption{Total numbers of emldetect detections with the 2-D and
  default PSFs, and the percentage change in numbers, for various
  `problem' cases, and for the `random' and the `clean' samples (see
  text). Negative percentage changes indicate that the 2-D PSF
  performs better at avoiding likely spurious sources in problem
  fields.}
\label{sourcesearch}
\begin{center}
\begin{tabular}{llrrc} \hline \hline
Set & Problem & N     & N         & (2-D $-$ def)/def \\ 
    & type    & (2-D) & (def) & perc. change \\ \hline
Problem & non-piled-up  & 4286 & 4491 & -4.6\% \\
        & piled-up           & 1171 & 1255 & -6.7\% \\
        & off-axis sources   & 1372 & 1388 & -1.2\% \\ 
        & OOTs (piled-up)      &  132 &  135 & -2.2\% \\
        & extended sources  &  896 &  893 & +0.3\% \\
        & reflection arcs & 1194 & 1172 & +1.9\% \\ \hline
Problem & Total   & 9417 & 9700 & -2.9\% \\ \hline  
Random  & Total   & 6395 & 6524 & -2.0\% \\ \hline  
Clean   & Total   & 2570 & 2575 & -0.2\% \\ \hline \hline
\end{tabular}
\end{center}
\end{table}

\begin{figure*}
\begin{center}
\includegraphics[width=7.4cm,angle=270]{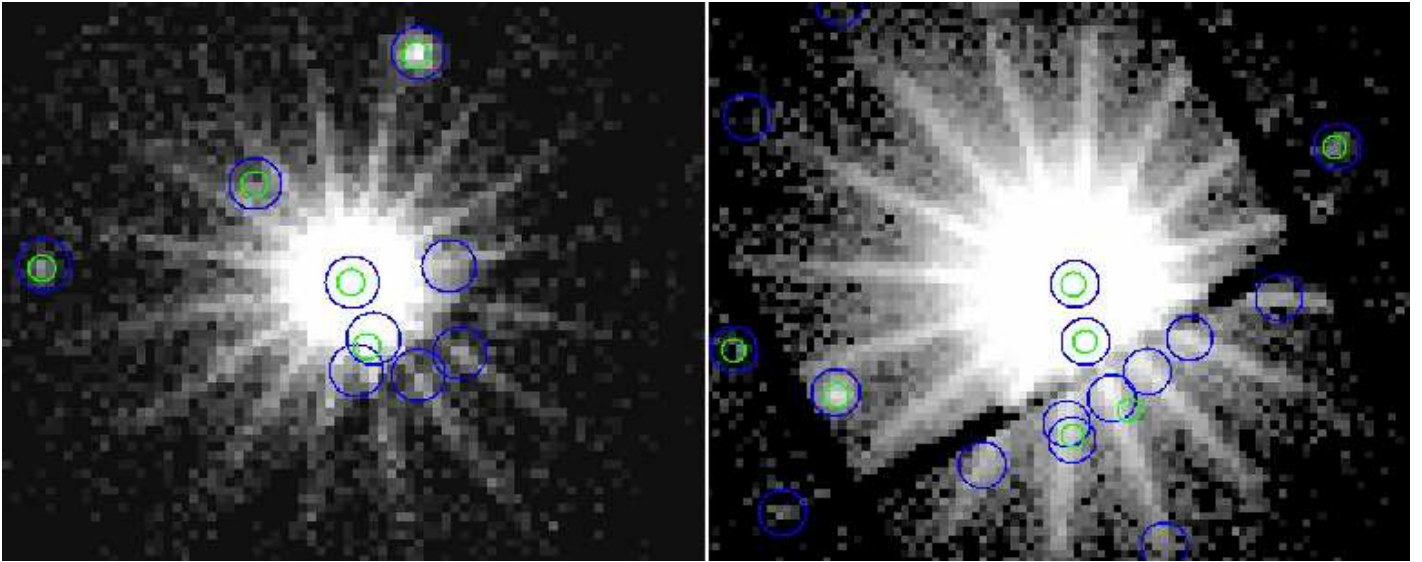}
\end{center}
\caption{Example output from a full all-EPIC source-detection analysis
  of ObsIDs 0107660201 (left) and 0302850201 (right). The blue circles
  show the sources detected using the default PSF and the yellow
  circles show the sources detected using the 2-D PSF. Many spurious
  sources previously detected by the default PSF in the spokes of the
  central bright source are now not detected by the 2-D PSF. }
\label{sourcefig}
\end{figure*}

Overall for the problem cases, the number of detections with the 2-D
PSF is lower (by 2.9\%) than for the default PSF. This is dominated by
the large reductions in the on-axis
non-piled-up and piled-up cases. The same
analysis applied to the random cases (where many similar `problems'
still exist) yields a smaller reduction than for the problem cases,
while for the clean cases, there is almost no change. This all
supports the idea that the 2-D PSF performs better at avoiding likely
spurious sources in problem fields. In fact, if one removes the clean
subset from the random set, the excess of detections by the default
PSF in the remaining `non-clean' cases is 3.1\%, consistent with the
`problem' 2.9\% in Table~\ref{sourcesearch}. Two examples showing the
improvement in the source detection around bright sources are shown in
Fig.\ref{sourcefig}.

We then focussed on the detection statistics around the (typically
brighter) objects that give rise to the excessive source detections.
Using the set of problem fields, source counts were recorded for 3
circular apertures centred on the problem source of interest, with
radii of 1\arcm, 2\arcm\ and a radius determined by eye that enclosed
the visible spoke structure. For insignificantly piled-up sources
(examining 53 fields, with essentially one bright object per field) we
find that use of the 2-D PSF reduces the number of detections around
bright sources by $\approx$23\% on average, and is not noticably
dependent on the aperture. Applying the same analysis to significantly
piled up sources (14 sources) and to bright off-axis sources (21
sources) yields changes of between 26\% and 31\%.

To examine more closely the distribution of sources around problem
objects, a visual analysis was conducted on a subset of objects,
counting sources that were either associated with the spoke features,
or suspect objects found elsewhere in the 2\arcm\ aperture but
generally close to the core of the source. The primary source
detection itself was excluded. This is inevitably a subjective
analysis and does not take account of the dependence on detection
likelihood, but it provides a useful assessment of the impact of the
2-D PSF. We find that for sources detected on spokes, overall the 2-D
PSF finds $\gtsim$42\% fewer detections than the default PSF (or
$\gtsim$92\% fewer if we consider only detections on spokes that are
not common to both PSFs). Likewise, for sources near the core of the
primary source, the 2-D PSF yields at least 22\% fewer detections, or
$\gtsim$69\% fewer if we exclude the common detections.

Thus, we conclude that use of the 2-D PSF substantially reduces the
number of detections compared to the default PSF and that this
reduction is primarily due to fewer detections on the spokes and in
the core regions of bright sources. These are precisely the components
of the PSF which are being better modelled by the 2-D PSF. These model
improvements suppress data-minus-model residuals that have hitherto
given rise to spurious detections when using the default PSF.

\subsection{Astrometry issues}

\begin{figure*}
\begin{center}
\includegraphics[width=9cm,angle=0]{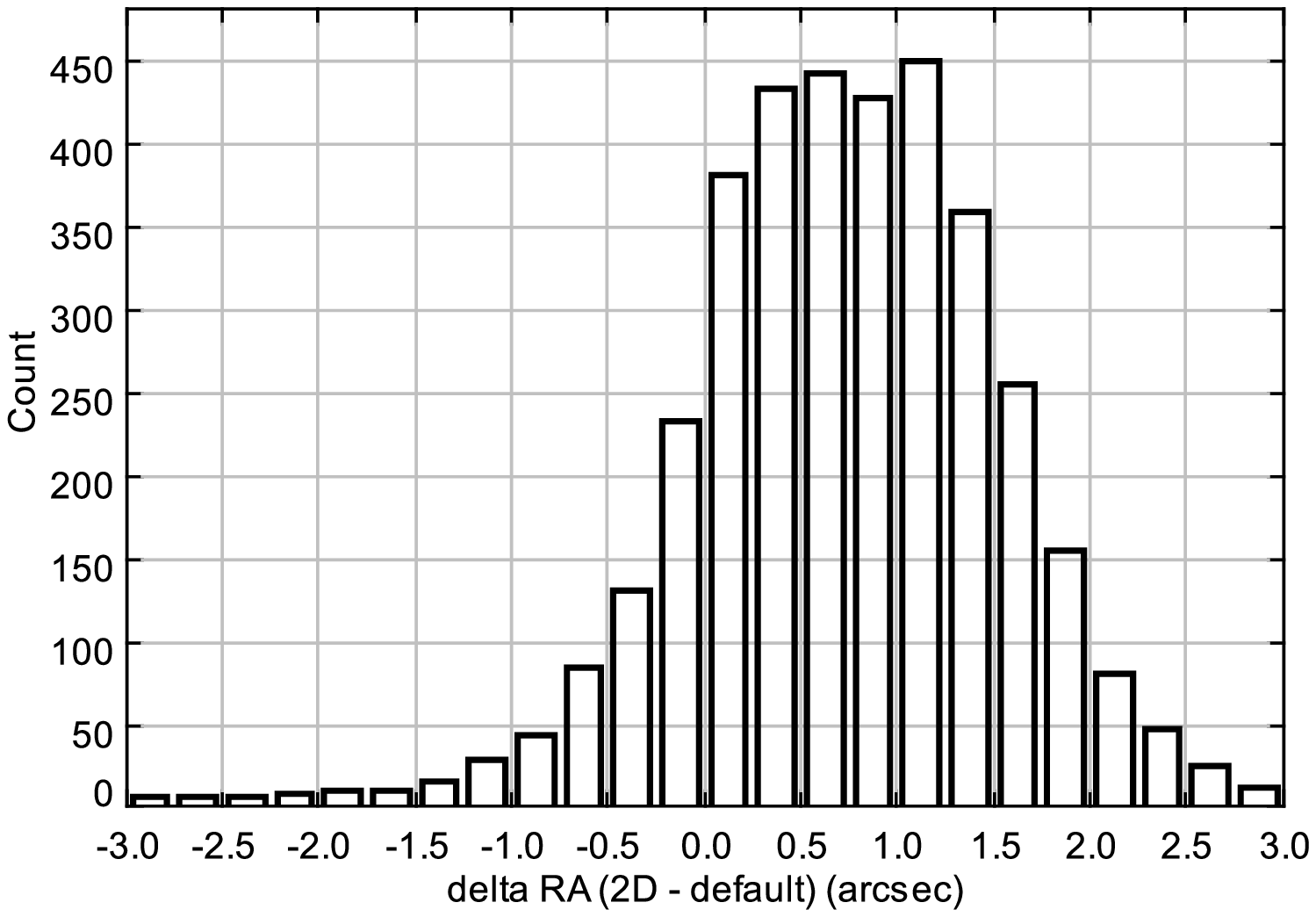}\includegraphics[width=9cm,angle=0]{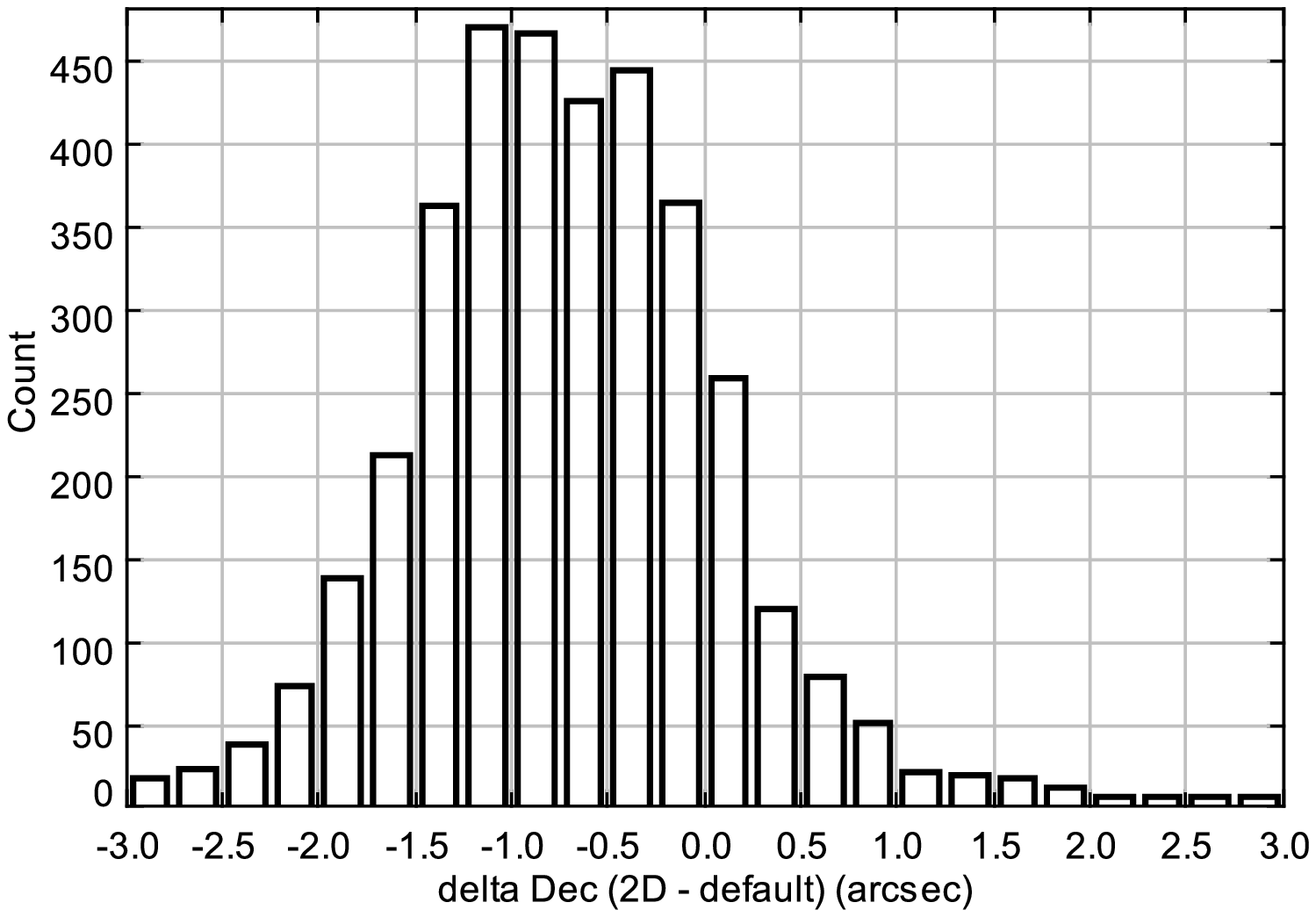}
\end{center}
\caption{Distributions of the (2D$-$default) change in returned (left) RA and
  (right) Dec of detected sources using the 2D (Ellbeta) and default
  (Medium) PSF for a sample of 70 clean observations.}
\label{shifts}
\end{figure*}

A major outstanding concern in the comparison of the 2-D PSF output
with the default PSF output was that large and significant positional
shifts are observed. Mean differences of $\approx$+0.8\arcs in RA and
$\approx$-0.8\arcs in Dec (2-D values minus default values) are
observed between the 2-D and default PSF results on any large set of
observations. Fig.\ref{shifts} shows the distributions of the RA and
Dec (2-D$-$default) offsets for a sample of 70 observations
(incorporating and extending on the clean sample of the previous
section).

\begin{figure*}
\begin{center}
\includegraphics[bb=23 23 454 398, width=9cm,angle=0]{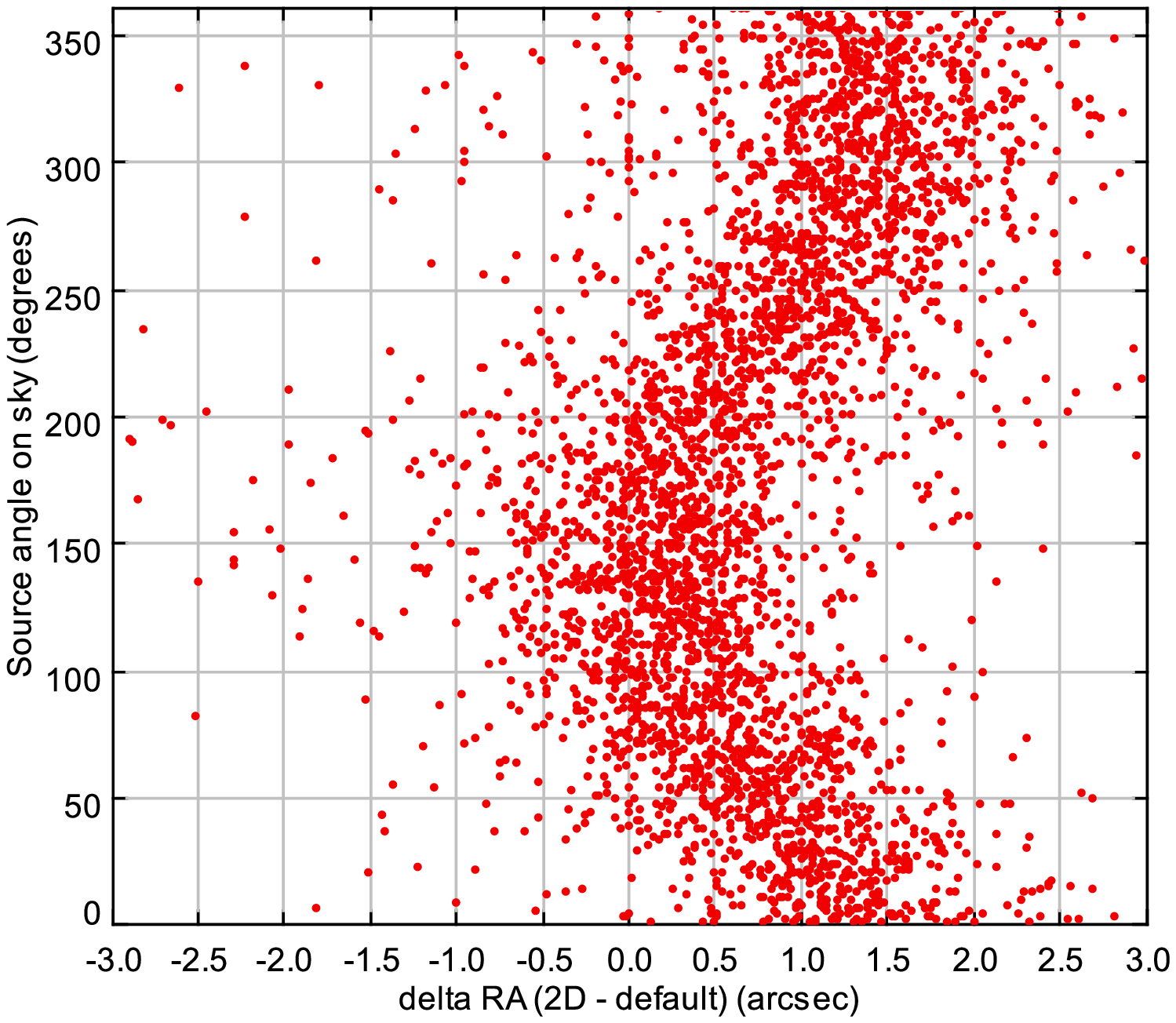}\includegraphics[bb=23 23 454 398, width=9cm,angle=0]{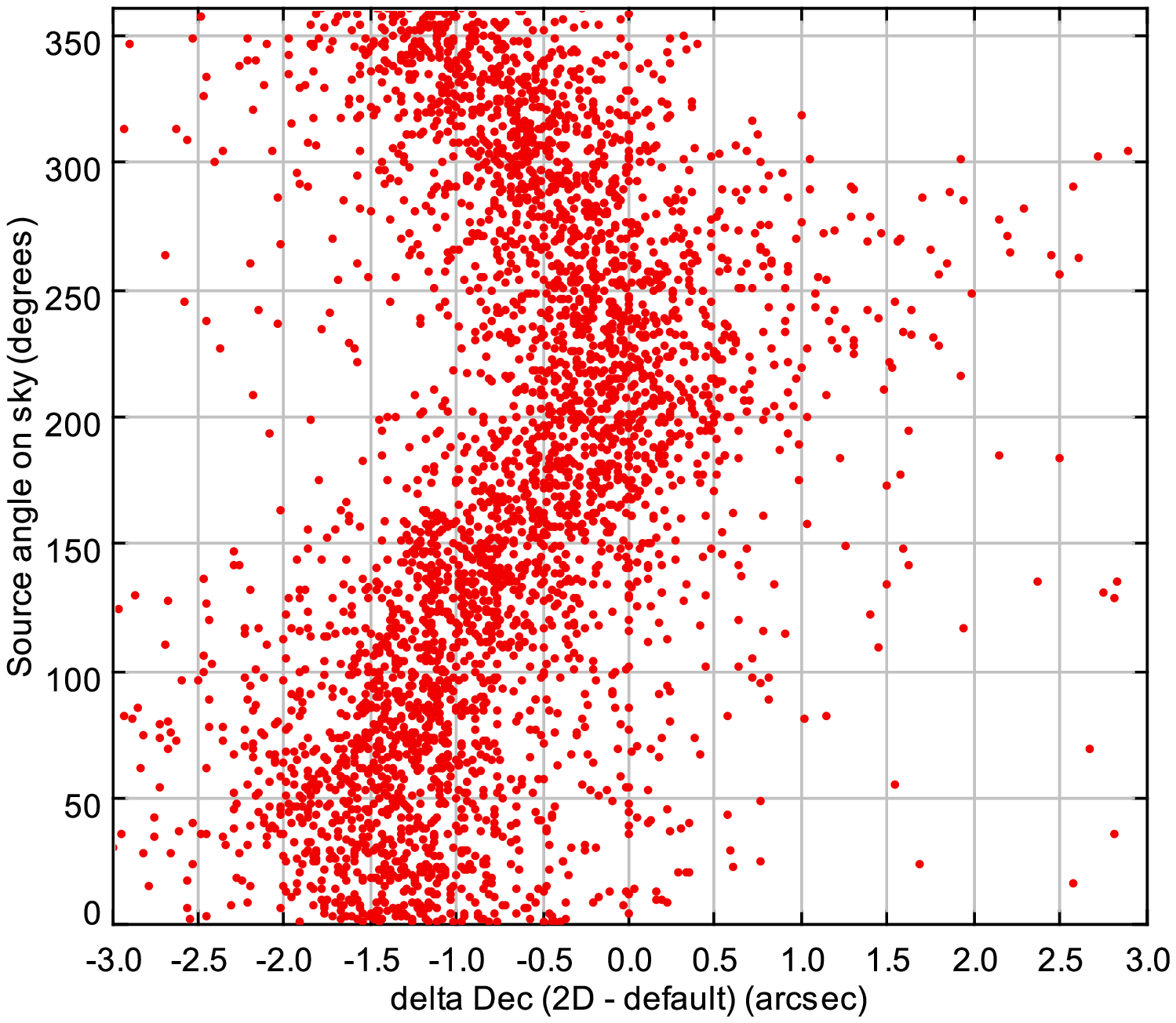}
\end{center}
\caption{(2D$-$default) offsets in (left) RA and (right) Dec of
  detected sources using the 2D (Ellbeta) and default (Medium) PSF
  plotted against the source angle on the sky (anti-clockwise from
  north) for the sample of 70 clean observations as in
  Fig.\ref{shifts}.}
\label{sinusoids}
\end{figure*}

It was not known whether this shift was due to the 2-D PSF or to the
default PSF (or to both). One source of confusion and possible error
is the fact that the default PSF (as described briefly in Sect.1.1)
consists of a set of images of dimensions 512$\times$512 and pixel
size 1.1\arcs$\times$1.1\arcs. It was noted that the assumption of
where the centre of one of these images is, and the propagation,
correctly or incorrectly, of this assumption through the relevant SAS
subsystems (PSF and image generation, PSF rotation, source-searching
etc.) could lead to possible errors. This thinking also indicated that
the rotation of the PSF could be vital in pinning down the root of the
positional offset problem, and indeed, when the (2-D $-$ default)
changes in RA and Dec are plotted against the source angle on the sky
(Fig.\ref{sinusoids}), the +0.8\arcs\ and -0.8\arcs\ offsets are
resolved into not only offset, but sinusoidal variations, indicating
that (i) not only is there an offset problem between the two PSFs, but
(ii) the rotation of one (or both) of the PSFs contains an additional
systematic sinusoidal error.

In order to evaluate which of the PSFs is at fault, it is necessary to
cross-correlate the X-ray positions obtained with the two PSFs with
good-quality optical positions. Here we need to introduce the
boresight matrix. This was calculated from fields of many bright
sources, e.g. OMC2/3 (Kirsch 2004), and subsequently refined using 430
individual fields (Altieri 2004), cross-correlating the source X-ray
positions (note, obtained using the default PSF) with their optical
positions, and observing which particular offset in RA, Dec and
position angle is seen to significantly optimize the correlation. This
final calibrated boresight misalignment matrix has been applied to the
determination of the X-ray sky position of every X-ray event in every
calibrated event file.

It is therefore the case that any translation or offset problem that
is due to the default (Medium) PSF is `corrected for' and calibrated
out when implementing the current boresight misalignment matrix. As
such, it is expected that the X-ray$-$optical source positional
offsets for the default PSF will be centred around zero, and indeed,
this is seen to be the case. The 2-D PSF offsets are furthermore seen
to be centred around the observed (Figs.\ref{shifts} \&
\ref{sinusoids}) +0.8\arcs\ and -0.8\arcs\ offsets.

The clean samples of source detections (using both the 2-D PSF and
the default PSF) were cross-correlated with the Sloan Digital Sky
Survey Quasar Catalog (Schneider \etal\ 2007) to a match radius of
10\arcs. In almost every case where there was a match, it was the only
match. We looked at the cross-correlations using the original
(emldetect-obtained) X-ray positions and using the X-ray positions
obtained by the SAS program eposcorr. This program, eposcorr, performs
a similar task to the boresight misalignment matrix calculation in
that, for each observation, it correlates the input (emldetect) source
positions with the positions from optical source catalogues, and
checks whether there are offsets in RA, Dec and position angle which
optimize the correlation. If there are optimum offsets, these are then
used to correct the input source positions which are then added as
separate columns to the input X-ray source list.

\begin{figure*}
\begin{center}
\includegraphics[width=8.2cm,angle=0]{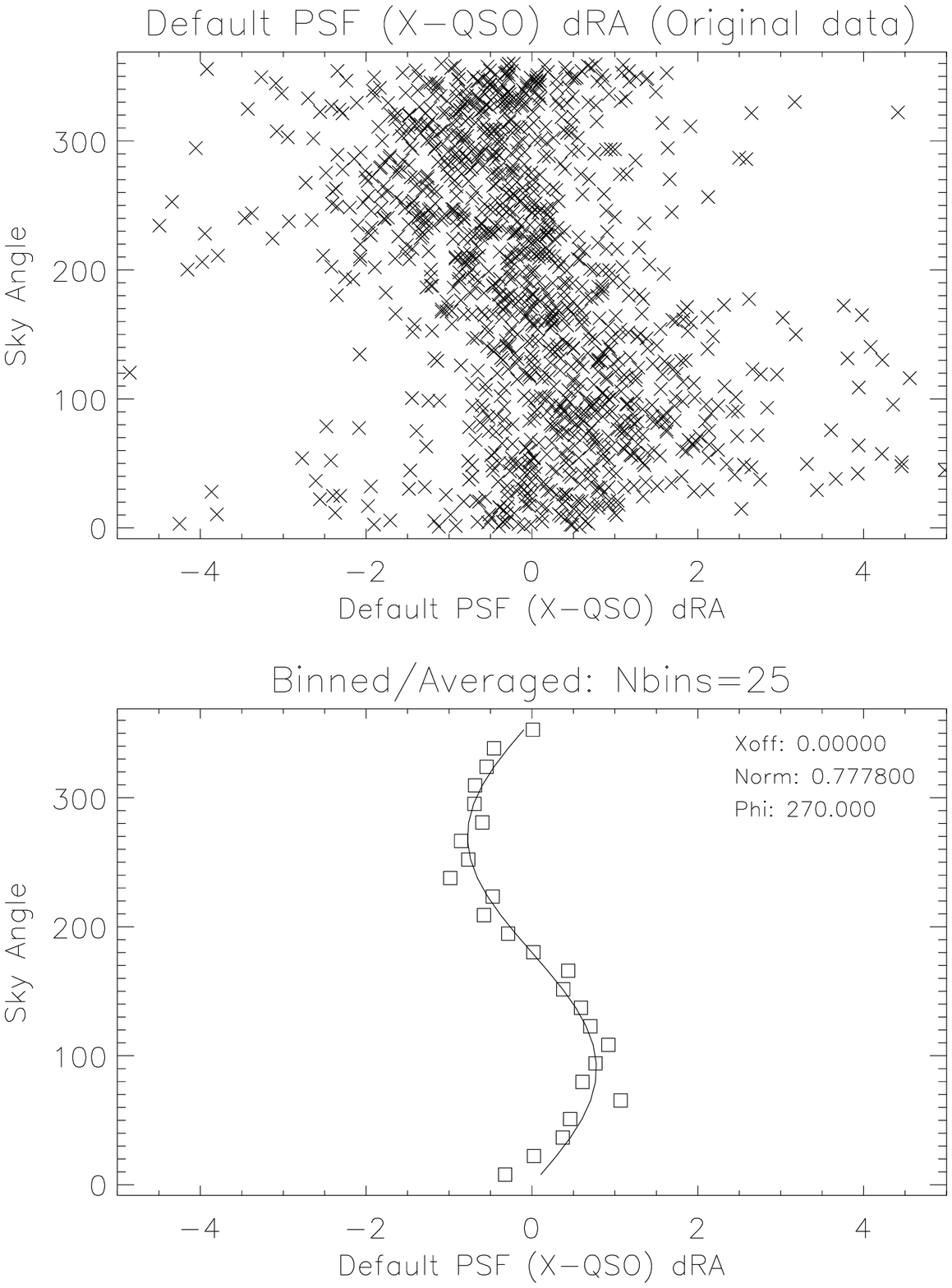}
\includegraphics[width=8.2cm,angle=0]{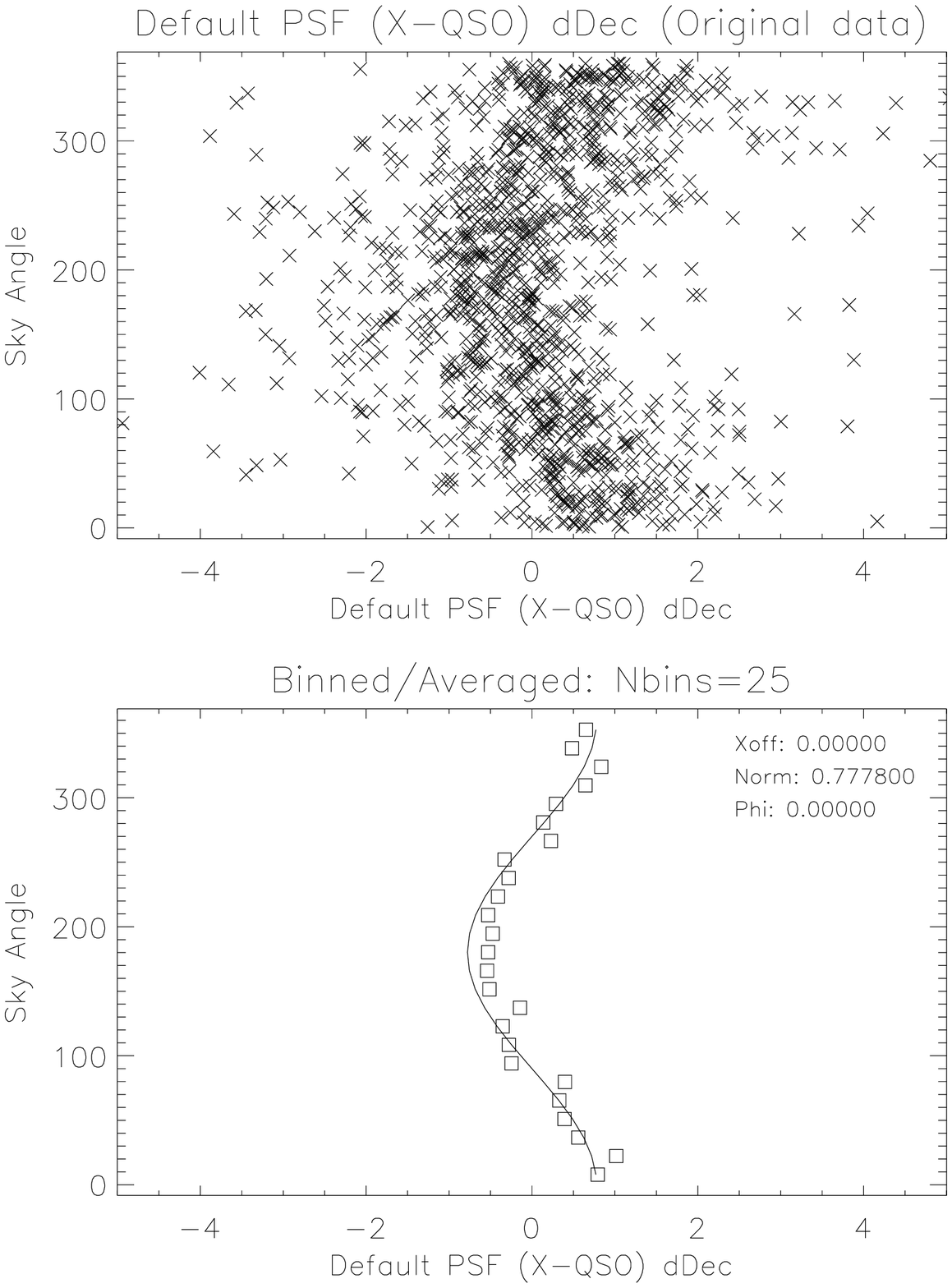}
\end{center}
\caption{(Top) X-ray position $-$ QSO optical position offsets in
  (left) RA and (right) Dec of default (Medium) PSF-detected X-ray
  sources with a QSO match, plotted against the source angle on the
  sky (anti-clockwise from north). X-ray positions are after
  eposcorr (see text). (Bottom) The same data, binned into 25 equi-angular bins
  and averaged.}
\label{xqsodef}
\end{figure*}

\begin{figure*}
\begin{center}
\includegraphics[width=8.2cm,angle=0]{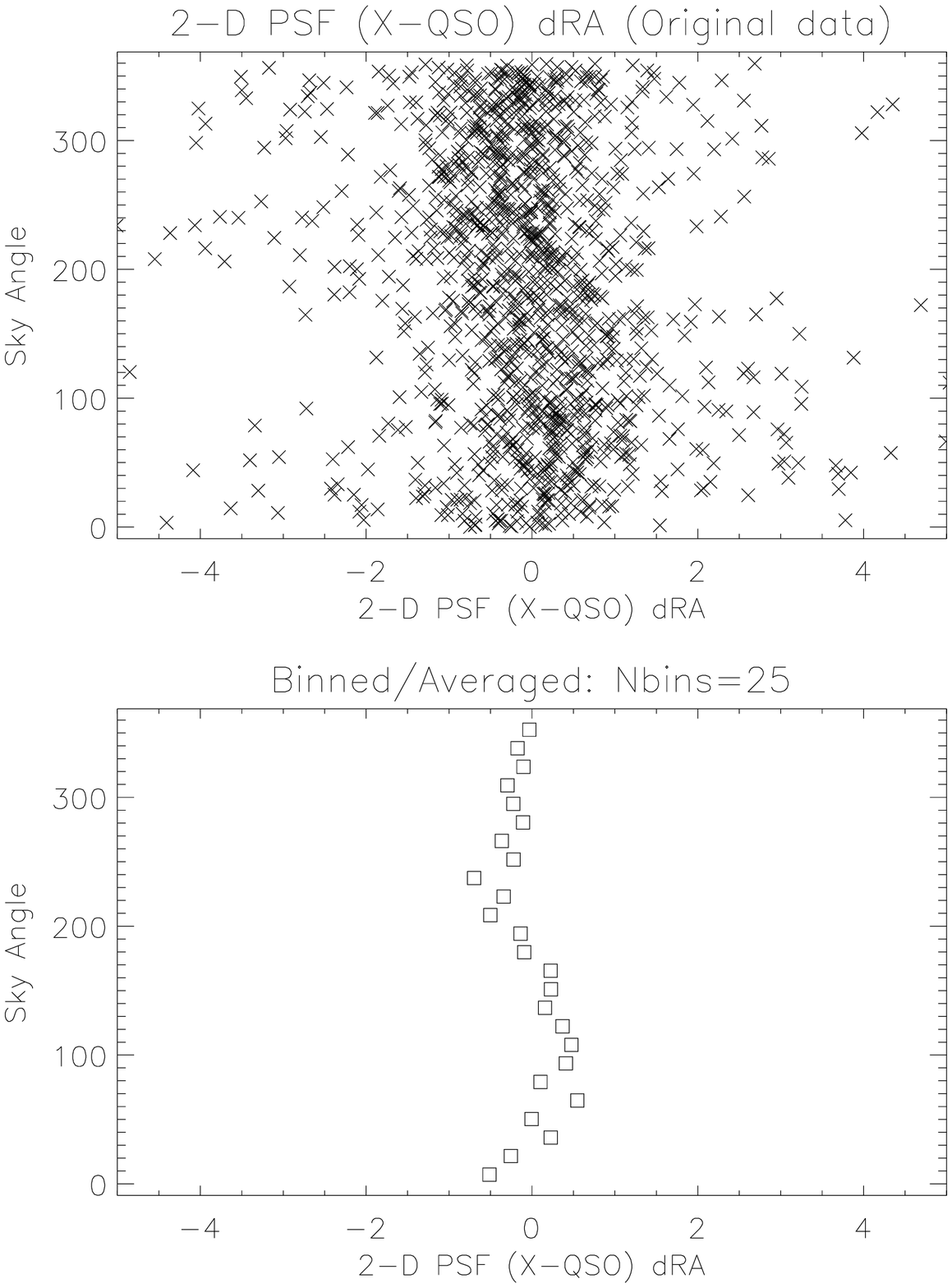}
\includegraphics[width=8.2cm,angle=0]{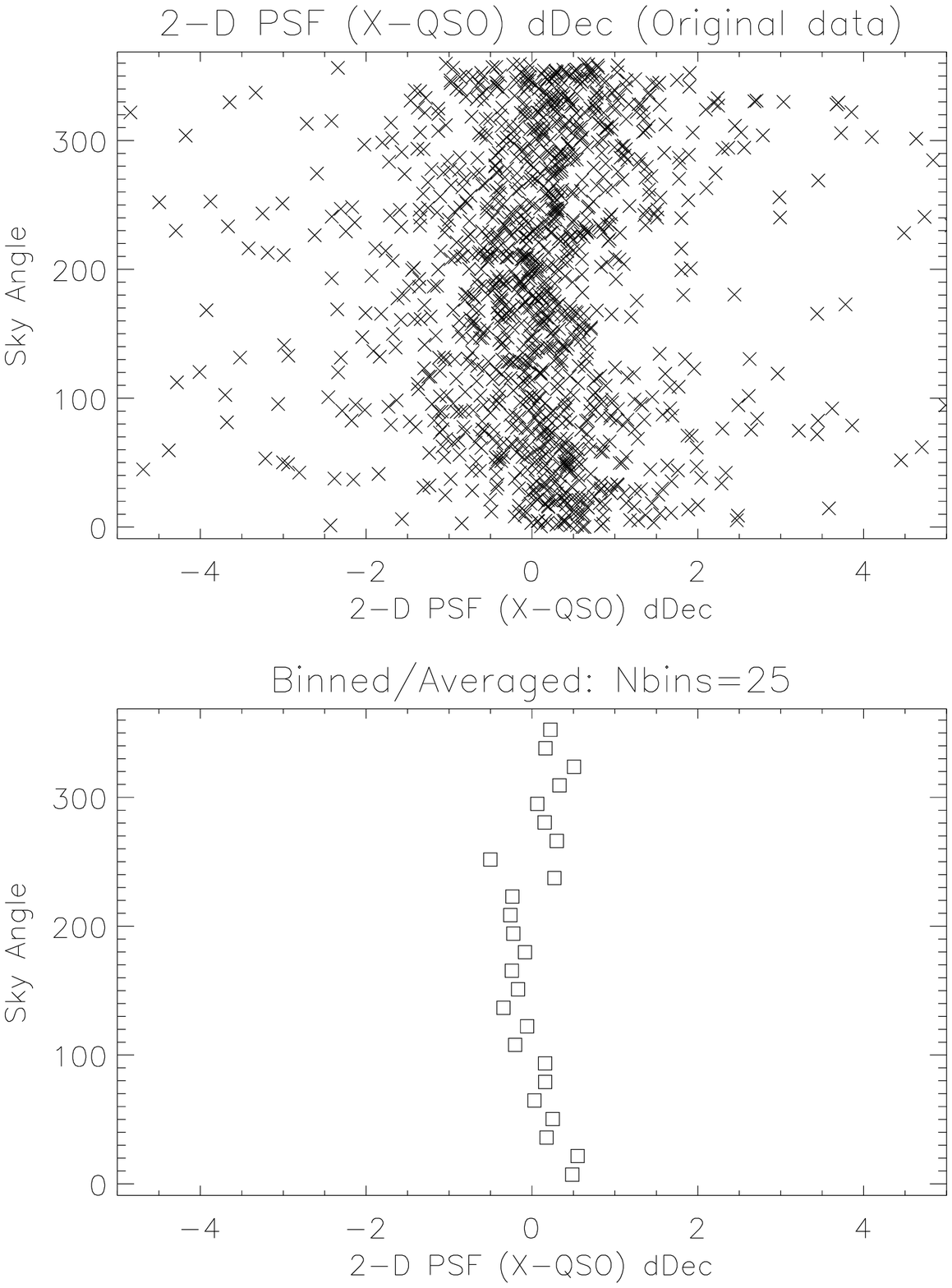}
\end{center}
\caption{(Top) X-ray position $-$ QSO optical position offsets in
  (left) RA and (right) Dec of 2-D (Ellbeta) PSF-detected X-ray
  sources with a QSO match, plotted against the source angle on the
  sky (anti-clockwise from north). X-ray positions are after
  eposcorr (see text). (Bottom) The same data, binned into 25 equi-angular bins
  and averaged.}
\label{xqso2d}
\end{figure*}

When the (X $-$ QSO) differences in RA and Dec between the X-ray
positions and the optical quasar positions are plotted against source
angle on the sky for both the default PSF (Fig.\ref{xqsodef}) and the
2-D PSF (Fig.\ref{xqso2d}, both figures using the eposcorr-corrected
output), several things are evident.

Firstly, regarding the offset problem, though the corresponding
figures prior to eposcorr (not shown) do indeed show the default PSF
distribution to be centred around zero offset, and the 2-D PSF
distribution to be centred around the +0.8\arcs\ and
-0.8\arcs\ offsets, the 2-D PSF eposcorr output (Fig.\ref{xqso2d}) is
centred around zero $-$ {\em i.e.} when it runs successfully, eposcorr
is able to correct the offset problem. Secondly, regarding the
sinusoidal problem, it is observed that it is for the default PSF case
(Fig.\ref{xqsodef}) that the large sinusoidal variations are seen. For
the 2-D PSF (Fig.\ref{xqso2d}) the variations are very much smaller.

Looking in more detail, it was suggested earlier that the sinusoidal
problem could be related to the default PSF CCF images, and the
assumptions regarding where the centres of these images are, and how
these assumptions are propagated through the entire PSF generation and
source-searching system. This now does appear to be the case, and
added to the lower panels of Fig.\ref{xqsodef} are simple calculated
cosine curves of what one would expect to see, given the simple offset
of half the diagonal of a single default PSF image pixel, rotated
around the 360 degrees of the detector. The data is seen to match this
simple model extremely well. It is therefore undoubtedly the default
PSF that is the cause of the sinusoid effect, and this problem has
existed for the entirety of the XMM-Newton mission. The full
positional capabilities of EPIC therefore have not been used to this
date. The improvement here is such that, selecting the 67\% of sources
with a very good X-ray positional error (centroid error $<$1\arcs),
the mean X-ray-QSO positional offset is reduced from
1.13\arcs\ (default) to 0.94\arcs\ (2-D), and the percentage of these
sources with an X-ray-QSO positional offset less than
1\arcs\ increases from 52\% (default) to 68\% (2-D). The percentage of
cases where it is the 2-D PSF X-ray position that is the closest to
the QSO position is 70\%.

Although the large sinusoids present in Fig.\ref{xqsodef} have been
largely removed in Fig.\ref{xqso2d}, they do not appear to have been
completely removed (and are at the 0.2\arcs$-$0.3\arcs\ level). One
should bear in mind that, if there is any inaccuracy at all present in
the true positioning of the PSF centre, and where it is assumed to be
in the PSF system, then a very similar sinusoidal variation will be
seen with source angle on the sky. The fact that some residual
curvature possibly exists in Fig.\ref{xqso2d} may indicate that the
situation is not quite perfect, and there may still be a small
residual misalignment of the centering of the PSF system and the true
PSF system.

At present, the 2-D PSF only returns the optimum positions when
eposcorr is run. Eposcorr only provides gross field shifts and cannot
correct for the residual sinusoidal effects. Also, the eposcorr task
can not be run on every EPIC dataset, for instance in cases where
there are very few (or very faint) X-ray sources in the field. In
order for the full positional improvements of the 2-D PSF to be usable
therefore, a revised boresight misalignment matrix first needs to be
calibrated, tested and incorporated correctly into the SAS (related
SAS changes may also be required). This work, building on the above
tests, is currently underway, and a future version of the SAS will
include these improvements in the positional accuracy of EPIC.

\section{Conclusions}

A new and fully comprehensive full-field-of-view (FOV) 2-D model of
the point spread functions (PSFs) of the three XMM-Newton EPIC
telescopes has been constructed. This has been performed via the
stacking, and bringing-together to a common reference frame, of a
large number of good quality, long-exposure, non piled-up, bright 
point sources from different positions within the full FOV of each
EPIC detector. The resultant general PSF envelopes were then
azimuthally filtered to construct the primary and secondary spoke
structures (plus their radial dependencies) and the large-scale gross
azimuthal PSF deformations that are observed. The PSF model also
includes an additional Gaussian core, which accounts for (at most) 2\%
of the enclosed energy flux in the EPIC-MOS cameras.

This PSF model is available for use within the XMM-Newton Science
Analysis System (SAS) via the usage of Current Calibration Files
(CCFs) XRT$i$\_XPSF\_0011.CCF and later versions. The modular nature
of the new PSF system allows for further corrections and refinements
in the future.

Initial EPIC source-searching tests using this new PSF model indicate
that it performs significantly better with regard to the major problem
with the previous PSF; that of large numbers of spurious source being
detected in the wings of, or close to bright point sources. The numbers
of these spurious sources detected with the new PSF model are greatly
reduced. 

These tests also uncovered a systematic error in the previous PSF
system, such that returned source RA and Dec values were seen to vary
sinusoidally about the true position with source azimuthal position
angle on the sky. This error in the previous PSF system (the amplitude
of the sinusoid being $\approx$0.8\arcs\ in RA and in Dec) has existed
since the beginning of the XMM-Newton mission, and affects all the
EPIC positional determinations performed thus far. Usage of the new
PSF results in a much smaller amplitude sinusoid, and therefore an
improved positional accuracy. SAS changes, including a revised
boresight misalignment matrix (presently under construction) are
required to make full use of the improvements in the positional
accuracy of EPIC introduced by the new PSF.

\begin{acknowledgements}

  The XMM-Newton project is an ESA Science Mission with instruments
  and contributions directly funded by ESA Member States and the USA
  (NASA). We thank Matteo Guainazzi and Steve Sembay for careful
  readings of the paper, and the referee for useful comments which
  have improved the paper. AMR and SRR acknowledge the support of
  STFC/UKSA/ESA funding.

\end{acknowledgements}


\end{document}